\documentclass[iop,twocolappendix,appendixfloats]{emulateapj}

\usepackage{graphicx}
\usepackage{amssymb}
\usepackage{multirow}
\usepackage{float}
\usepackage{comment}

\newcommand{\be}{\begin{equation}}
\newcommand{\ee}{\end{equation}}

\newcommand{\ba}{\begin{eqnarray}}
\newcommand{\ea}{\end{eqnarray}}

\begin{document}
 
\title{{\em Herschel} and {\em Spitzer} observations of slowly rotating, nearby isolated neutron stars}
\author{B. Posselt}
\affil{Department of Astronomy \& Astrophysics, Pennsylvania State University, 525 Davey Lab,University Park, PA 16802, USA }
\email{posselt@psu.edu}
\author{G. G. Pavlov}
\affil{Department of Astronomy \& Astrophysics, Pennsylvania State University, 525 Davey Lab,University Park, PA 16802, USA}
\author{S. Popov}
\affil{Sternberg Astronomical Institute, Lomonosov Moscow State University, Moscow 119992, Russia}
\author{S. Wachter}
\affil{Max Planck Institute for Astronomy, K\"onigsstuhl 17, 69117 Heidelberg, Germany}

\begin{abstract}
Supernova fallback disks around neutron stars have been discussed to influence the evolution of the diverse neutron star populations. Slowly rotating neutron stars are most promising to find such disks.
Searching for the cold and warm debris of old fallback disks, we carried out {\em Herschel} PACS ($70$\,$\mu$m, $160$\,$\mu$m) and  {\em Spitzer} IRAC ($3.6$\,$\mu$m, $4.5$\,$\mu$m) observations of eight slowly rotating ($P\approx 3 - 11$\,s) nearby ($<1$\,kpc) isolated neutron stars. 
{\em Herschel} detected $160$\,$\mu$m emission ($>5\sigma$) at locations consistent with the positions of the neutron stars RX\,J0806.4--4123 and RX\,J2143.0+0654. 
No other significant infrared emission was detected from the eight neutron stars.
We estimate probabilities of 63\%, 33\% and 3\%  that, respectively, none, one, or both {\em Herschel} PACS $160$\,$\mu$m detections are unrelated excess sources due to background source confusion or an interstellar cirrus.
If the $160$\,$\mu$m emission is indeed related to cold (10\,K to 22\,K) dust around the neutron stars, this dust is absorbing and re-emitting $\sim 10$\% to $\sim 20$\% of the neutron stars' X-rays. 
Such high efficiencies would be at least three orders of magnitude larger than the efficiencies of debris disks around nondegenerate stars.
While thin dusty disks around the neutron stars can be excluded as counterparts of the $160$\,$\mu$m emission, dusty asteroid belts constitute a viable option.
\end{abstract}


\section{Introduction}
During the last decade sensitive X-ray observations have revealed a rather surprising diversity of neutron stars (NSs); for a review see, e.g., \citet{Kaspi2010}. 
While most of the currently known $\sim 2000$ NSs are rotation-powered radio pulsars, there are also rotating radio transients, central compact objects (CCOs) in supernova remnants, X-ray thermal isolated NSs (XTINSs) and magnetars. The group of the magnetars, consisting of the Anomalous X-ray Pulsars (AXPs) and the Soft Gamma-ray Repeaters (SGRs), posed particular challenges to the simplistic NS model as a rotating magnetic dipole which converts rotational energy into electromagnetic pulsar emission. 
Magnetar quiescent emission and the occasionally observed huge magnetar flares require an additional energy source. 
The prevailing model for magnetars is that of young, isolated NSs powered by decay of very large magnetic fields  \citep{Duncan1992}. Within this model the flares are understood as reconnection events in a twisted magnetosphere (e.g., \citealt{Thompson2002}). 
In addition to the dipole magnetic fields, recent magneto-thermal NS evolution models consider crustal toroidal magnetic fields as well (e.g., \citealt{Pons2007,Pons2009}). These models have been successful in explaining, for example, the observed numbers of magnetars, XTINSs and radio pulsars \citep{Popov2010}, or the observed properties of CCOs \citep{Vigano2012}.\\

However, there are viable additional or alternative factors which may influence the evolution and observed diversity of NSs. Different masses of supernova fallback disks could exert different disk torques, leading to different accretion rates and ultimately to the diversity of observed NS classes (e.g., \citealt{Alpar2001, Alpar2001b}).  
Explaining the emission of magnetars as the most extreme case has become the litmus test for this theory. Accretion disks around magnetars have initially been suggested and discussed by \citet{Paradijs1995} and \citet{Chatterjee2000}. The theory was supported by recent works (e.g., \citealt{Truemper2013, Ertan2006}) which explain the X-ray spectral and timing properties, as well as the radio, optical and IR properties of magnetars. Note, however, that these models cannot produce the observed giant magnetar flares which are attributed to local ultra-strong multipole fields in an analogy to star spots \citep{Truemper2013}. Recently, \citet{Calicskan2013} expanded the application of the fallback model to a different NS class, represented by radio pulsars with high magnetic fields. 
These recent results imply that fallback disks currently cannot be excluded as crucial ingredients in the NS evolution. 
In fact, such disks are a general prediction of supernova models \citep{Michel1981,Chevalier1989}, however, observationally they have remained rather elusive; for a review about searches for disks around NSs see \citet{Wang2014}.
There is so far only one example of a likely fallback disk around the 3.9\,kpc distant AXP 4U 0142+61, detected with {\em Spitzer} IRAC \citep{Wang2006, Ertan2007}. 
According to \citet{Eksi2005}, fallback disks are expected to be rare because they are likely to be disrupted when the newly born NS spins rapidly through the propeller stage, at which inflowing matter
would be expelled instead of being accreted. The fallback disks can survive if the initial NS spin is slow enough ($P_0 \ge 40$\,ms at a magnetic moment of $\mu = 10^{30}$\,G\,cm$^{3}$).
The presence of disks around highly magnetized NSs today is therefore, in principle,
related to the initial birth parameters of these NSs,
in particular, to the origin of the strong fields in highly magnetized NSs.
It is important to obtain more comprehensive observational constraints on the presence of supernova fallback disks around the different NS classes. Slowly rotating NSs, such as the magnetars and the XTINS, appear to be most promising to find such disks.

\section{Sample selection and working hypotheses}
In the framework of the fallback disk model, \citet{Alpar2001,Alpar2007} proposed that different initial disk masses lead to the different NS classes. In this picture, magnetars are expected to have the most massive disks, followed by those of the XTINS, while rotation powered radio pulsars have negligible or no disks.
\citet{Truemper2010} suggested that XTINSs may be more evolved versions of AXPs with fallback disks.
It is interesting to note that in the $P-\dot{P}$ diagram the XTINS are located together with a few high-$B$ pulsars between the separated groups of magnetars and radio pulsars (see, e.g., Figure\,2 in \citealt{Kaspi2010}). Thus, these NSs  may constitute a key population to unify the different NS classes.\\
 
Of the 26 currently known magnetars and magnetar candidates only 5 have been investigated at infrared wavelengths  \citep{Olausen2013}. Apart from the detections of AXP 4U 0142+61 \citep{Wang2006} and 1E 2259+586 \citep{Kaplan2009}, the shallow mid-IR {\sl Spitzer} searches of three other magnetars were not very constraining \citep{Wang2007a}. It is difficult to constrain the disk emission around magnetars because most of these NSs are far away. The median distance of magnetars with known distances is 8.5\,kpc \citep{Olausen2013}\footnote{McGill SGR/AXP Online Catalog (March 2014): www.physics.mcgill.ca/\~pulsar/magnetar/main.html}. 
The XTINS, however, are all expected to be within 1\,kpc based on parallaxes and the interstellar absorption of their soft X-ray spectra (e.g., \citealt{Kaplan2009, Posselt2007}). Such distances enable constraining searches for fallback disks for the whole known XTINS population. We selected all known slowly rotating ($P>3$\,s) NSs within  1\,kpc. Our sample includes all the seven known XTINS and the nearby high$-B$ radio pulsar PSR\,J1848--1952 which has similar spin-down properties (e.g., \citealt{Kaplan2009}).\\

In the framework of the fallback disk model, potential XTINS disks have likely smaller masses than magnetar disks, are likely older than those, and are likely passive, i.e., without significant on-going accretion on the NS. The latter property can be concluded from the very stable X-ray emission of the XTINSs with the notable exception of RX\,J0720.4--3125 (e.g., \citealt{Haberl2007,Kerkwijk2007, Hohle2012}). \\

The composition and evolution of fallback disks are unknown. 
Dust grain evolution was shown to decouple from gas evolution in nearly all (protoplanetary) disk configurations if the initial dust-to-gas ratio is larger than 0.03 \citep{Pinte2014}. 
Given the supernova origin, one can reasonably assume metal-rich disk material, i.e., a decoupled dust evolution in the disk. Gas is easier removed from the disks than dust grains by the pulsar wind and the interstellar ram pressure. We will therefore assume dust-rich disks as a working hypothesis for potential residual XTINS fallback disks. In comparison to other stars, the potential XTINS disks may more likely resemble transitional or debris disks -- such as those around main-sequence stars (e.g., \citealt{Wyatt2008} and references therein), central stars in planetary nebulae (e.g., \citealt{Su2007}), or white dwarfs (e.g, \citealt{Debes2012} and references therein) -- than protostellar/protoplanetary disks around young stars.

\section{Observations}
\label{obs}
\subsection{{\em Herschel} Observations}
{\em Herschel} observations of eight isolated NSs were carried out over a period of $\sim 1$\,year using the Photodetector Array Camera and Spectrometer (PACS, \citealt{Poglitsch2010}). 
We obtained the images in mini-map  mode in the blue (60--85\,$\mu$m) and red (130--210\,$\mu$m) bands with the respective PACS bolometer arrays operated in parallel. The mini-map mode is a particular case of the scan-map mode, where 
two observations (scan and cross-scan) are carried out which have scan angles in array coordinates along the two array diagonal directions, 70 and 110 degrees.
An individual scan-map observation has a pattern of parallel, overlapping scan lines connected by turnaround loops. The mini-map mode provides a good point source sensitivity with a  relatively large homogeneous coverage (about $50\arcsec$ in diameter) \footnote{PACS Observer's Manual; Chapter 5.2; herschel.esac.esa.int/Docs/PACS/html/pacs\_om.html}.
Our observations were done with medium scan speed ($20\arcsec$\,s$^{-1}$), the mapping parameters were 10 scan legs with scan lengths of $3\arcmin$ with a separation of $4\arcsec$ between them. A repetition factor of 12 was chosen to reach point source sensitivities close to the expected confusion noise (for a details about the confusion noise in the PACS bands and its estimation see, e.g., the  Herschel Observers' Manual\footnote{herschel.esac.esa.int/Docs/Herschel/html/Observatory.html}, Chapter 4.3. and the HERSCHEL-HSC-DOC-0886\footnote{herschel.esac.esa.int/Docs/HCNE/pdf/HCNE\_ScienceDoc.pdf} document by C. Kiss).
Our observations are listed in Table~\ref{herschelobs}. Overall, each Astronomical Observation Request (AOR)
was 7\,ks.
The beam sizes (full width of half maximum, FWHM) of the blue and red band at medium speed are $5\farcs{5} \times 5\farcs{8}$ and $10\farcs{5} \times 12\farcs{02}$, respectively \citep{Poglitsch2010}. 

\begin{deluxetable}{lccc}
\tablecaption{List of Herschel observations \label{herschelobs}}
\tablehead{
\colhead{Object} & \colhead{Date} & \colhead{ObsIDs} & \colhead{PointAcc\tablenotemark{a}}} 
\startdata
RX\,J0420.0--5022 & 2012-01-07 & 1342236938/9 & $1\farcs{1}$ \\
RX\,J0720.4--3125 & 2011-11-10 & 1342232218/9 & $1\farcs{1}$ \\
RX\,J0806.4--4123 & 2011-10-29 & 1342231572/3 & $1\farcs{1}$ \\
RX\,J1308.6+2127  & 2011-07-11 & 1342223951/2 & $1\farcs{5}$ \\
RX\,J1605.3+3249  & 2011-07-20 & 1342224486/7 & $1\farcs{5}$ \\
RX\,J1856.5--3754 & 2011-03-15 & 1342216058/9 & $2\farcs{4}$ \\
PSR\,J1848--1952  & 2012-03-13 & 1342241353/4 & $1\farcs{1}$ \\
RX\,J2143.0+0654  & 2012-04-23 & 1342244870/1 & $1\farcs{1}$ 
\enddata
\tablenotetext{a}{$1\sigma$ absolute pointing accuracy; see text in Section~\ref{herscheldatared}.}
\end{deluxetable}

\subsection{{\em Spitzer} Observations}
We observed RX\,J0720.4--3125 with the {\em Spitzer} Space Telescope\footnote{irsa.ipac.caltech.edu/data/SPITZER/docs/spitzermission/} \citep{Werner2004} during the post-cryogenic mission in January 2011 (PI: Posselt). 
The observations were done with the Infrared Array Camera (IRAC; \citealt{Fazio2004}) at $3.6$\,$\mu$m and $4.5$\,$\mu$m. We used a frame time of 100\,s and 31 dither positions in each band.  
We also consider here some of the NS observations of another {\em Spitzer} program (PI: Wachter) which searched for fallback disks around a large number of nearby pulsars using the IRAC $4.5$\,$\mu$m channel. Those observations were done from September 2011 to December 2012. In each observation a 100\,s frame time and 30 dither positions were used.
The {\em Spitzer} observations discussed in this paper are listed in Table~\ref{spitzerobs}.
The IRAC field of view is $5\farcm{2} \times 5\farcm{2}$, the native IRAC pixel size is $1\farcs{2}$, the FWHM of point spread function at $3.6$\,$\mu$m and $4.5$\,$\mu$m is $1\farcs{66}$ and $1\farcs{72}$, respectively \citep{Fazio2004}.\\

\begin{table}[h]
\begin{center}
\caption{List of Spitzer observations \label{spitzerobs}}
\begin{tabular}{lccc}
\tableline\tableline
Object & Date  & Channels & AOR\\
\tableline\tableline
RX\,J0420 & 2011-09-17 & $4.5$\,$\mu$m & 42478336\\
RX\,J0720 & 2011-11-21 & $3.6$\,\&\,$4.5$\,$\mu$m & 39947264\\
RX\,J0806 & 2012-04-30 & $4.5$\,$\mu$m & 42478592\\
RX\,J1308 & 2011-07-24 & $4.5$\,$\mu$m & 42478848\\
RX\,J1605 & 2012-05-01 & $4.5$\,$\mu$m & 42479616\\
RX\,J1856 & 2011-11-12 & $4.5$\,$\mu$m & 42479360\\
RX\,J2143 & 2012-12-31 & $4.5$\,$\mu$m & 42479872 \\
\tableline\tableline
\end{tabular}
\end{center}
\end{table}

\section{Data Reduction and flux measurement procedure}
\label{datared}
\subsection{{\em Herschel} Observations}
\label{herscheldatared}
Using the {\em Herschel} Interactive Processing Environment (HIPE), v9.0\footnote{herschel.esac.esa.int/HIPE\_download.shtml}, we reduced the data  based on the deep-survey scan-map pipeline scripts of the PACS photometer pipeline. The PACS calibration file set PACS\_CAL\_41\_0\footnote{pacs.ster.kuleuven.be/Data/pcal-community/release-note-v41.html} was applied.
We reprocessed the data from the Level 0 to the final maps (Level 2.5) applying the usual processing steps outlined in \citet{Poglitsch2010} and \citet{Lutz2011}.\\
A critical step in the data processing is the removal of the 1/f noise, the heavy flicker noise of the PACS readout system. In the case of a point source observation, a high pass filter along the time line of the observation allows one to remove most of the 1/f noise, which increases the sensitivity of the observation. The effect of the high-pass filter on the PACS point spread function and the noise, the resulting flux removal and mask strategies to avoid the former are discussed in detail by \citet{Popesso2012}. 
According to this study, the high-pass filter most efficiently removes noise in case of high data redundancy, i.e., large repetition factors. Flux losses of point sources are reduced best by putting circular mask patches on prior source positions. The filter width and patch sizes need to be carefully chosen since their influences dominate the global and local noise in the PACS maps. In our data processing we checked maps produced with a masking technique where only sources above a certain threshold are masked before the combined map is produced. This method can be expected to have the highest sensitivity for faint point source identification. However, not all observations show a PACS source near the isolated neutron star (INS) position. Aiming to reduce flux losses in our flux (limit) estimates, we therefore masked a region with radius $10\arcsec$ around each expected INS position for our final map productions.\\    
The pixel noise of individual pixels in a PACS photometer map is correlated due to the 1/f noise and due to the applied \texttt{drizzle} projection method by \citet{Fruchter2002} during the map creation process. \citet{Popesso2012} calculated the individual noise contributions in dependence of the high-pass filter width, the output pixel size and the pixel drop fraction used in the \texttt{drizzle} method. 
In general, the noise level is lower in the high data redundancy case compared to the low data redundancy case, and smaller pixel sizes and smaller pixel drop fractions can be chosen in the former case to achieve the same noise limit.
We checked different sets of parameters. As a good final choice for all our maps, we used  pixel sizes of $1\arcsec$ in the blue band,  $2\arcsec$ in the red band, a pixel drop fraction of 0.04 in both bands and high-pass filter widths of 15 readouts in the blue band and 25 readouts in the red band.\\

The expected positions of the INSs at the time of the {\em Herschel} observations were calculated from positions and proper motions (if known) in the literature, and they are listed in Table~\ref{position}.
Position errors are usually below $1\arcsec$. We comment on individual INS uncertainties in Section~\ref{results} and Appendix \ref{detresults} if there are close neighbor sources in the  {\em Herschel} images. The {\em Herschel} absolute pointing error ranges from $1\sigma \approx 2\farcs{4}$ for our earliest {\em Herschel} observation (RX\,J1856) to $1\sigma \approx 1\farcs{1}$ for our latest {\em Herschel} observations according to the  {\em Herschel} calibration and HIPE teams\footnote{herschel.esac.esa.int/twiki/bin/view/Public/SummaryPointing}. The pointing errors are listed in Table~\ref{herschelobs}.\\

\begin{deluxetable*}{lccccccccc}
\tablecaption{Positions and proper motions of the eight neutron stars \label{position}}
\tablewidth{0pt}
\tablehead{
\colhead{Object} & \colhead{R.A.}  & \colhead{DEC} & \colhead{PosErr\tablenotemark{a}} &\colhead{Epoch} & \colhead{Ref.} & \colhead{$\mu_{\alpha} \cos \delta$\tablenotemark{a}} & \colhead{$\mu_{\delta}$\tablenotemark{a}} & \colhead{Ref.} & \colhead{PosErrH\tablenotemark{b}}\\ 
\colhead{} & \colhead{h m s}  & \colhead{d m s} & \colhead{$\arcsec$} & \colhead{MJD} & \colhead{\tablenotemark{c}} & \colhead{mas yr$^{-1}$} & \colhead{mas yr$^{-1}$} & \colhead{\tablenotemark{c}} & \colhead{$\arcsec$ (90\%)}}\\
\startdata
RX\,J0420  & 04 20 01.94 & $-50$ 22 47.8 & $1.1$ (90\%) & 54122 & 1 & \multicolumn{2}{c}{$\mu <123 $ $(2\sigma)$} & 2 & $1.2$ \\
RX\,J0720  & 07 20 24.93 & $-31$ 25 49.78 & $0.02,0.02$ ($1\sigma$) & 54477 & 3 & $-93 \pm 5$\,($1\sigma$)& $49 \pm 5$\,($1\sigma$) & 3 &  0.06 \\
RX\,J0806  & 08 06 23.40 & $-41$ 22 30.9 & $0.6$ (90\%) & 52326 &4 & \multicolumn{2}{c}{$\mu < 86$ $(2\sigma)$} & 2 & $0.9$ \\
RX\,J1308 & 13 08 48.27 & $+21$ 27 06.8 & 0.6\,(90\%) & 51719 & 5 & $-207 \pm 20$\,($2\sigma$) & $84 \pm 20$\,($2\sigma$) & 2 & 0.7 \\
RX\,J1605  & 16 05 18.52 & $+32$ 49 18.0 & $0.3,0.3$ ($1\sigma$) & 52111 &  6 & $25 \pm 16$\,(90\%) & $142 \pm 15$\,(90\%) & 7 & 0.7\\
RX\,J1856  & 18 56 35.795 & $-37$ 54 35.54 & 0.1\,(90\%)\tablenotemark{d} & 52868 & 8 & $326.6 \pm 0.5$\,($1\sigma$) & $-61.9 \pm 0.4$\,($1\sigma$) & 8 & 0.1\\
PSR\,J1848 & 18 48 18.04 & $-19$ 52 31 & 0.6, 7\,($2\sigma$) & 48695 & 9 & $39 \pm 157$\,($2\sigma$)  & $-1200 \pm 1900$\,($2\sigma$) & 9 & 3.1,9\tablenotemark{e}  \\
RX\,J2143  & 21 43 03.4 & $+06$ 54 17.5  & 0.2,0.2\,($1\sigma$) & 54388 & 10 & \multicolumn{2}{c}{$\mu <700$ (90\%)} & tw\tablenotemark{f} & $3.2$ 
\enddata
\tablenotetext{a}{Uncertainty levels as quoted in the respective papers. One value instead of two indicates  the radial error.}
\tablenotetext{b}{Positional radial error (90\% confidence level) of the NS position at the epoch of the respective {\em Herschel} observation, calculated using the Rayleigh distribution. In the case of different (position, proper motion) error values in R.A. and DEC we use the larger error for our estimate.}
\tablenotetext{c}{References: 1 -- \citet{Mignani2009}, 2 -- \citet{Motch2009}, 3 -- \citet{Eisenbeiss2010}, 4 -- \citet{Haberl2004}, 5 -- \citet{Kaplan2002}, 6 -- \citet{Kaplan2003}, 7 -- \citet{Motch2005}, 8 -- \citet{Walter2010}, 9 -- \citet{Hobbs2004}, 10 -- \citet{Schwope2009}.
 }
\tablenotetext{d}{Absolute positional error not listed in \citet{Walter2010}, we assume a reasonable value of $0\farcs{1}$ (90\% confidence)}
\tablenotetext{e}{Individual errors in R.A. and DEC, see discussion in Section~\ref{psr1848}}
\tablenotetext{f}{This work, see Section~\ref{sub2143}. }
\end{deluxetable*}

We use apertures with sky annuli to measure source fluxes and apply the corresponding aperture corrections using the respective HIPE tasks. 
The calibration study by \citet{Popesso2012} provided a method to calculate aperture flux errors from the coverage of individual map pixels by considering the chosen data reduction parameters. 
First, the coverage-error map relation as well as the cross-correlation factor  are calculated in dependence on the chosen pixel size, pixel drop fraction and high-pass filter width.  
Then, the coverage is used to calculate the error of the individual map pixels. 
The pixel errors are quadratically added for the respective aperture size.
Finally, the cross-correlation factor is applied.
This cross-correlation correction factor accounts for both components in the correlation noise, the projection itself and the residual $1/f$ noise not removed by the high-pass filter. 
Since we give aperture-corrected fluxes, we apply the aperture correction factor to the estimated error as well.
We checked whether the determined final error was of the same order as the variance of background apertures in nearby source-free regions and found this to be the case.
We list $5\sigma$ errors for the aperture-corrected fluxes in Table~\ref{herschelres}.\\
 
If there is no source at the INS position, we consider several apertures to measure the standard deviation of the background in the central, source-free region of the map (close to the INS) where the exposure coverage is reasonably homogeneous. Depending on how crowded the field is, we usually obtain results from 7 to 15 apertures with radii of $5\arcsec$ or $7\arcsec$.
We list $5\sigma$ values for the flux limits in Table~\ref{herschelres}. \\ 

We neglect the PACS photometer color corrections (for estimates of the monochromatic flux densities for different spectral shapes) because they are close to one for a wide range of expected disk temperatures (30\,K to 1000\,K). However, if highly accurate fluxes for a particular disk model are desired, we refer to the respective {\em Herschel} PACS Technical Note PICC-ME-TN-038\footnote{herschel.esac.esa.int/twiki/pub/Public/PacsCalibrationWeb\protect\\/cc\_report\_v1.pdf}. 
 
\subsection{{\em Spitzer} Observations}
\label{spitzerdatared}
For the data reduction of the {\em Spitzer} IRAC data, we used {\em Spitzer}'s MOsaicker and Point source Extractor package, MOPEX\footnote{irsa.ipac.caltech.edu/data/SPITZER/docs/dataanalysistools\protect\\/tools/mopex/}, v.18.5.6.  
In the following we give an exemplary description of the data reduction for the IRAC data of RX\,J0720.4--3125 (two bands). The data reduction for the $4.5$\,$\mu$m data of the other INSs is done following the same procedures.\\

In each case, we start from the artifact-corrected basic calibrated data (cBCD) frames.  We removed one cBCD frame from the $3.6$\,$\mu$m data set of RX\,J0720.4--3125 because it had a bad column at the target position. From the respective $4.5$\,$\mu$m data set we removed 5 frames due to artifacts, possibly the so-called column pull-up, at or close to the source position.
Using the \texttt{mosaic} task we obtained a combined image with a pixel size of $0\farcs{6}$. 
We checked the alignment of the IRAC mosaic astrometry with the 2MASS Point Source Catalog \citep{Skrutskie2006} using 114 2MASS sources with the highest quality flag AAA in the field.
The astrometric accuracy of the $3.6$\,$\mu$m and the $4.5$\,$\mu$m mosaics of RX\,J0720.4--3125 are $0\farcs{27}$ and $0\farcs{28}$, respectively.
Similar astrometric accuracies are found for the other INS {\em Spitzer} mosaics.
The expected INS positions were calculated for the times of the respective {\em Spitzer} observations using known positions and proper motions as described for {\em Herschel} in Section~\ref{herscheldatared}.\\

The IRAC Point Response Function (PRF) is essentially the convolution of a box, having the size of the image pixel, with the point spread function (e.g., MOPEX user guide, chapter 8.9\footnote{irsa.ipac.caltech.edu/data/SPITZER/docs/dataanalysistools\protect\\/tools/mopex/mopexusersguide/89/\#\_Toc320000083}. To consider a PRF which is variable in the field of view, one uses the so-called PRF maps which are provided by the Spitzer Science Center\footnote{irsa.ipac.caltech.edu/data/SPITZER/docs/irac\protect\\/calibrationfiles/psfprf/} for the IRAC channels.
The PRF is required for point source fitting photometry, which we performed with the \texttt{apex multi-frame} task on each individual cBCD frame. Applying then the \texttt{apex-qa} task, we subtracted the fitted point sources from the individual cBCD frame to obtain a residual mosaic. Such residual mosaics are useful if neighboring sources contaminate the flux at the source position.\\

We use aperture photometry to determine IRAC fluxes and flux limits.
We apply aperture radii of 2 native IRAC pixels, corresponding to 4 mosaic pixels, and aperture correction factors of 1.21 and 1.22 for the $3.6$\,$\mu$m and the $4.5$\,$\mu$m measurements, respectively \citep{Reach2005}. 
To obtain an upper limit, we determine not only the source aperture flux, $F_{\rm SRC}$, but also the background level, $F_{\rm BG}$, and the standard deviation, $\sigma$, of several (usually 8-10) source-free apertures close to the source position. If necessary and possible, we use the residual mosaics for these measurements. 
In the case of a non-detection, we define the upper flux limit $F_{5\sigma}=F_{\rm SRC}-F_{\rm BG}+5\sigma$. \\ 

The derived IRAC fluxes are in Jy and correspond nominally to those fluxes one expects to measure in the respective IRAC channel for a source with $F_{\nu} \propto \nu^{-1}$. Sources with different spectral shapes, and in particular, very red sources require a color correction. The IRAC Instrument Handbook\footnote{irsa.ipac.caltech.edu/data/SPITZER/docs/irac\protect\\/iracinstrumenthandbook/18/} (section 4.4) lists color corrections for different power law indices and blackbody temperatures. 
In Sections~\ref{sub0806}, \ref{sub2143} and \ref{sub0420} to \ref{psr1848}, we give the aperture-corrected flux values without color correction, while in Table~\ref{herschelres} we exemplarily apply the color corrections for a $T=200$\,K blackbody as an approximation for the spectral shape of potential warm disk emission.

\subsection{Complementary {\em WISE} data}
\label{wise}
The {\em Wide-field Infrared Survey Explorer} ({\em WISE}, \citealt{Wright2010}) mapped the sky  in 2010 at 3.4 (W1), 4.6(W2), 12(W3), and 22 (W4) $\mu$m with an angular resolution of $6\farcs{1}$, $6\farcs{4}$, $6\farcs{5}$, and $12\arcsec$, respectively. We use the all-sky atlas images and the {\em WISE} Source Catalog to investigate the regions of our INSs. We cross-checked for astrometric shifts between the positions of known 2MASS sources or sources in our previous $H$-band VLT observations and source positions in {\em WISE} bands, as well as for noticeable astrometric shifts between {\em WISE} sources and the {\em Herschel} sources. 
While the astrometric calibrations of the first three {\em WISE} bands generally agree well with NIR astrometry, there are often noticeable shifts in the fourth band which are most easily seen when comparing the last two {\em WISE} bands. Astrometric biases in the {\em WISE} all-sky atlas images are discussed on the IPAC webpage\footnote{wise2.ipac.caltech.edu/docs/release/allsky/expsup/sec2\_3g.html}. While we suspect a shifted W4 astrometry for several of our fields, we do not attempt to improve the W4 astrometric calibration because the scarcity, faintness and spatial extent of the W4 sources hinder a good, unique astrometric calibration. We ignore the W4 band in the following.\\

We searched for potential {\em WISE} counterparts of the INSs or neighboring Herschel sources in the remaining bands and discuss our findings in the respective subsections.
In Table~\ref{WISElimits} we provide the {\em WISE} field limits measured in Vega magnitudes.
The limits correspond to the requirement that the IR sources have signal-to-noise ratios SNR$\ge 5$ in the respective band.  
To obtain these limits, we considered all {\em WISE} sources in a box area $1^{\circ} \times 1^{\circ} $ around the INS (usually around 20000 sources). Sources with SNR$=5$ usually have a magnitude uncertainty of $\approx 0.22$\,mag. \\

To convert the Vega magnitudes into flux density units such as Jy, one has to know and account for the source spectrum in the respective {\em WISE} band by applying a color correction. Due to the wide {\em WISE} bands, these color corrections can be very different for stars and much cooler blackbody-like sources (asteroids, disks); see, e.g., \citet{Cutri2012wise}, Section IV.4.h; \citet{Wright2010}. As an example, we give fluxes for a $T=200$\,K blackbody additionally to the Vega magnitudes in Table~\ref{WISElimits}, but caution that flux limits should in principle be more carefully calculated for any specific (disk) models under investigation.    

\section{Results}
\label{results}
There is no significant emission in the {\em WISE}, {\em Spitzer}, {\em Herschel} bands for six of the eight investigated NSs. In Sections~\ref{sub0806} and \ref{sub2143}, we present detailed results for the two NSs where significant {\em Herschel} red band emission was detected near the location of the NS. 
Appendix~\ref{detresults} includes finding charts and measurement details on the six NSs having only upper limits.  We summarize our {\em Herschel} and color-corrected {\em Spitzer} flux (limit) results in Table~\ref{herschelres}, and the {\em WISE} field magnitude limits in  Table~\ref{WISElimits}.

\begin{table}[t]
\begin{center}
\caption{{\em WISE} NS field limits \label{WISElimits}}
\begin{tabular}{lccc|ccc}
\tableline\tableline
Object & W1 & W2 & W3 & W1& W2 & W3 \\
&$3.4$\,$\mu$m & $4.6$\,$\mu$m & $12$\,$\mu$m& \multicolumn{3}{c}{for a BB with $T=200$K}\\
 & [mag] & [mag] &[mag] & [$\mu $Jy] &[$\mu $Jy] &[$\mu $Jy] \\
\tableline\tableline
RX\,J0420 &  18.4 &   17.0 &   12.8 &	 6   &     20  & 220	\\
RX\,J0720 &  17.2 &   15.7 &   11.4 &	 21  &     67  & 837	\\
RX\,J0806 &  17.5 &   16.3 &   12.2 &	 15  &     38  & 383	\\
RX\,J1308 &  17.3 &   15.9 &   11.7 &	 18  &     56  & 595	\\
RX\,J1605 &  17.7 &   16.2 &   12.1 &	 12  &     41  & 435	\\
RX\,J1856 &  17.0 &   15.5 &   11.2 &	 24  &     80  & 952	\\
PSR\,J1848  &  15.3 &   14.3 &   11.0 &	 110 &     240 & 1145	\\
RX\,J2143  &  17.3 &   15.8 &   11.6 &	 18  &     61  & 690	\\
\tableline\tableline
\end{tabular}
\end{center}
\tablecomments{Vega magnitude limits in the field of the NSs considering {\em WISE} detections with a SNR$>5$. The last three columns show the
color-corrected {\em WISE} fluxes if a 200\,K blackbody emitter is assumed for the spectral shape.}
\end{table}

\subsection{RX\,J0806.4--4123}
\label{sub0806}

\begin{figure}[h]
\begin{center}
{\includegraphics[width=85mm]{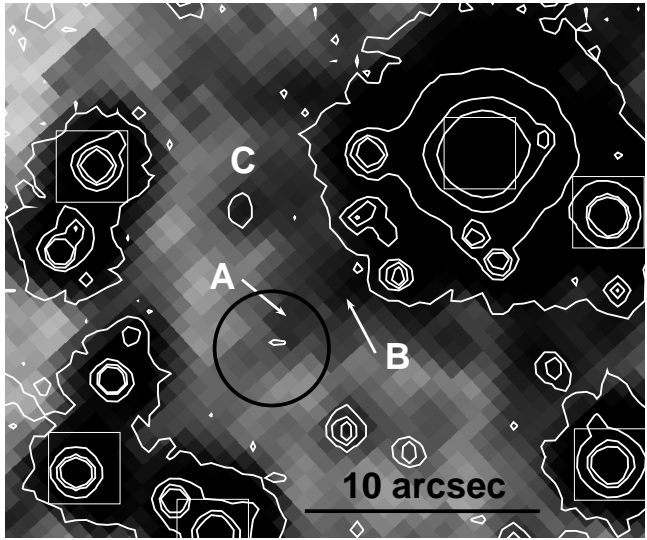}} \\
\caption{{\em Spitzer} IRAC2 ($4.5$\,$\mu$m) map around RX\,J0806.4--4123. The image is  $27\arcsec \times 23\arcsec$, North is up, East to the left. White contours are from our previous VLT $H$-band observations \citep{Posselt2009}. The white boxes indicate positions of 2MASS point sources, illustrating the good astrometric calibration. The INS position is marked with a black circle with a radius of 2 native IRAC pixels (4 image pixels, $2.4\arcsec$). Note that the astrometric uncertainty for the expected RX\,J0806.4--4123 position is less than $1\arcsec$.
The color scales of all Figures in this Section have been manually tweaked to emphasize faint fluxes in the surrounding of the NS position. None of these Figures is smoothed.
\label{j0806S}}
\end{center}
\end{figure}

The {\em Spitzer} IRAC2 ($4.5$\,$\mu$m) observation of RX\,J0806.4--4123 is well aligned with previous NIR observations (e.g, 2MASS point sources), having a $1\sigma$ astrometric accuracy of $0\farcs{2}$ (using 196 AAA 2MASS sources).
\citet{Posselt2009} investigated VLT ISAAC $H$-band images of this region. Figure~\ref{j0806S} shows that the major $H$-band sources correspond well to the {\em Spitzer} sources. 
The {\em Chandra} X-ray position of RX\,J0806.4--4123 has a positional uncertainty of $0\farcs{6}$ (90\% confidence, \citealt{Haberl2004}). Together with the upper limit on the INS proper motion at the time of the {\em Spitzer} observation ($0\farcs{88}$ ($2\sigma$), \citealt{Motch2009}), the overall positional uncertainty of the expected RX\,J0806.4--4123 position in IRAC2  is about $0\farcs{9}$ (90\% confidence level) in the worst case.\\

There appears to be faint emission in the $4.5$\,$\mu$m band within this range in the north-west direction of the nominal INS position, see source A in Figure~\ref{j0806S}. The emission is so faint that we were not able to detect it with the MOPEX PRF-fitting methods. It may be spurious emission; however, there are similarly faint sources which correspond to known NIR sources (e.g., source C in Figure~\ref{j0806S}).
We were not able to produce a reasonable \texttt{apex-qa} residual mosaic in which the brighter sources would be reliably subtracted. As the $H$-band contours show, the IRAC2 emission may contain several IR sources which are not recognized by the PRF-fit. For example, it would be useful to subtract source B, but this source is not even recognized as an individual source in our PRF fits. 
If we do flux aperture measurements at the INS position (with radius of 2 native IRAC pixels), we obtain nominally an aperture-corrected flux of $F^{4.5\mu \rm m}=5.2 \pm  1.9(1\sigma)$\,$\mu$Jy.  Since one can argue about the reliability of such $2.8\sigma$ emission, we also estimated the aperture-corrected upper flux limit $F^{4.5\mu \rm m}_{5\sigma}=14.7$\,$\mu$Jy. 
We note that there is a very faint NIR source at the INS position in the 2004 VLT $H$-band data too, but the NIR emission is below the $3\sigma$ significance level as well \citep{Posselt2009}.\\

\begin{figure}[h]
\begin{center}
{\includegraphics[width=85mm]{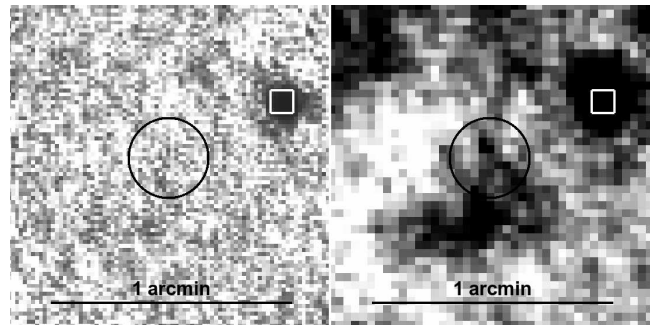}} \\
\caption{{\em Herschel} PACS maps around RX\,J0806.4--4123. The blue (60--85\,$\mu$m) and red (130--210\,$\mu$m) bands are shown in the left and right panels, respectively. Each map is $1\farcm{3} \times 1\farcm{3}$, North is up, East is to the left. The circle with a radius of $10\arcsec$ shows the NS  position. The white square marks a common source in {\em Herschel} PACS, {\em WISE} and VLT ISAAC images.
\label{j0806H}}
\end{center}
\end{figure}

\begin{figure}[h]
\begin{center}
{\includegraphics[width=85mm]{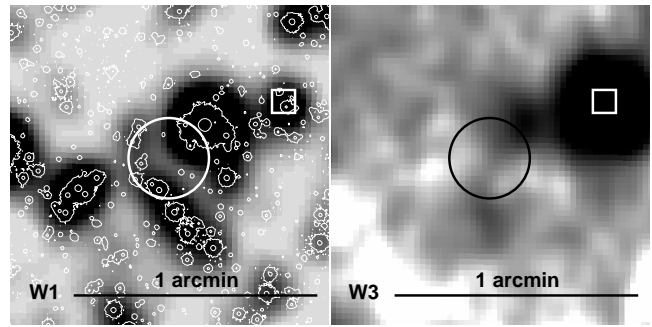}} \\
\caption{{\em WISE} W1 ($3.6$\,$\mu$m) and W3 (12\,$\mu$m) maps around the position of RX\,J0806.4--4123. The same field of view is shown as in the {\em Herschel} PACS maps of Figure~\ref{j0806H}.The white contours in the W1 image are from the VLT ISAAC $H$-band image \citep{Posselt2009}. The white square marks a common source in {\em Herschel} PACS, {\em WISE} and VLT ISAAC images.
\label{j0806W}}
\end{center}
\end{figure}

In the {\em Herschel} PACS blue band, there is no significant emission around the position of RX\,J0806.4--4123. Using apertures in different source-free regions and at the target position, we determined the $5\sigma$ upper flux limit for the INS as $F^{\rm blue}_{5\sigma} = 4.9$\,mJy. 
There is, however, emission around the INS position in the {\em Herschel} PACS red band, see Figure~\ref{j0806H}. We use the X-ray position of RX\,J0806.4--4123 from \citet{Haberl2004}. This position has an uncertainty of $0\farcs{6}$ (90\% confidence). In addition, an uncertainty of $0\farcs{83}$ needs to be considered due to the $2\sigma$ upper limit on the proper motion, $86$\,mas\,yr$^{-1}$. In the {\em Herschel} field of view, there is only one unambiguous point source in common with other infrared data, e.g., VLT $H$-band or {\em WISE}. The positions of this point source agree well. Comparing the extended emission with the {\em WISE} data, we conclude that the nominal  {\em Herschel} PACS pointing accuracy of $1\sigma \approx 1\farcs{1}$ can be assumed. The PACS red emission around the INS could either consist of several point sources or include extended emission. The peak closest to the INS has a spatial separation of $\approx 2\farcs{5}$ from the X-ray position.
We use a $5\arcsec$ aperture for the flux measurement and 10 other  apertures on source-free regions to measure the background. Using an aperture correction of $3.4$, we estimate a flux of $F^{\rm red} \approx 10 \pm 5 (5\sigma)$\,mJy at the position of the INS. In Section~\ref{alternatives} we discuss the likelihood of the {\em Herschel} PACS red band emission being not associated with the NS.\\    
We use the previous $H$-band image for cross-checks with the {\em WISE} W1-W3 data, see Figure~\ref{j0806W}. 
There is no noticeable astrometric shift between these bands. There is no apparent counterpart in the {\em WISE} data to the emission in the {\em Herschel} PACS red band. Higher spatial resolution is required to investigate the {\em Herschel} emission around the INS position.

\subsection{RX\,J2143.0+0654}
\label{sub2143} 
\begin{figure}[b]
\begin{center}
{\includegraphics[width=85mm]{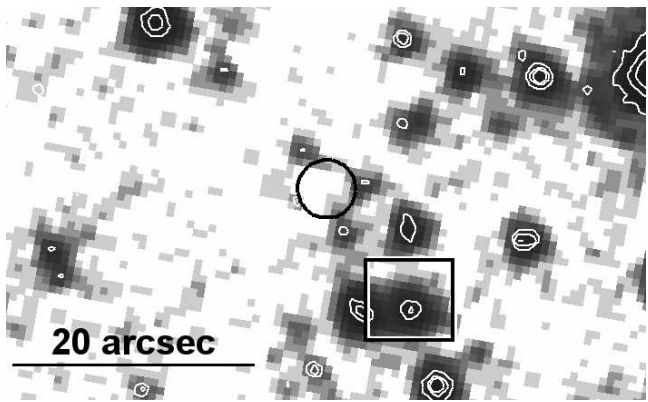}} \\
\caption{{\em Spitzer} IRAC2 ($4.5$\,$\mu$m) map around the position of RX\,J2143.0+0654. The white contour highlight sources in the $H$-band \citep{Posselt2009}. 
The black circle with a radius of $2\farcs{4}$ marks the INS position, the square marks a bright source common in the {\em Herschel} PACS, {\em WISE} W1 and W2, and $H$-bands. 
\label{j2143s}}
\end{center}
\end{figure}
The position of RX\,J2143.0+0654 is known with an accuracy of $0\farcs{2}$ from optical observations \citep{Schwope2009}, but its proper motion is unknown. Comparing the $Chandra$, XMM-$Newton$, and optical  positions from different epochs as shown by  \citet{Schwope2009}, one can infer that the proper motion must be smaller than 700\,mas\,yr$^{-1}$. Thus, at the time of {\em Herschel} and {\em Spitzer} observations, the INS could have moved from the optical position by $<3\farcs{2}$ and  $<3\farcs{7}$, respectively.\\ 

There is no emission at the position of RX\,J2143.0+0654 in the {\em Spitzer} IRAC $4.5\mu$m  mosaic image, see Figure~\ref{j2143s}. There are, however, four  {\em Spitzer} sources within $<3\farcs{7}$ of the nominal NS position. Since the same sources were detected as faint $H$-band sources in the 2003 VLT observations by \citet{Posselt2009}, none of them can be the NS counterpart.
The aperture corrected $5\sigma$ upper flux limit at the nominal NS position is $F^{4.5\mu \rm m}_{5\sigma}=3.0$\,$\mu$Jy.\\

\begin{figure}
\begin{center}
{\includegraphics[width=85mm]{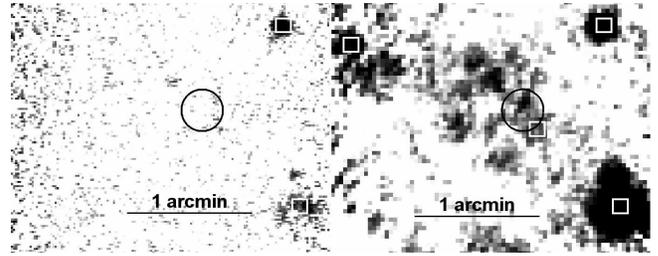}} \\
\caption{{\em Herschel} PACS maps around RX\,J2143.0+0654. The blue (60--85\,$\mu$m) and red (130--210\,$\mu$m) bands are shown in the left and right panels, respectively. Each map is $2\farcm{5} \times 2\arcmin$, North is up, East is to the left. The circle with a radius of $10\arcsec$ marks the INS position, the white squares mark bright sources common in the {\em Herschel} PACS, {\em WISE} W1 and W2, and $H$-bands. 
\label{j2143H}}
\end{center}
\end{figure}

\begin{figure}[t]
\begin{center}
{\includegraphics[width=85mm]{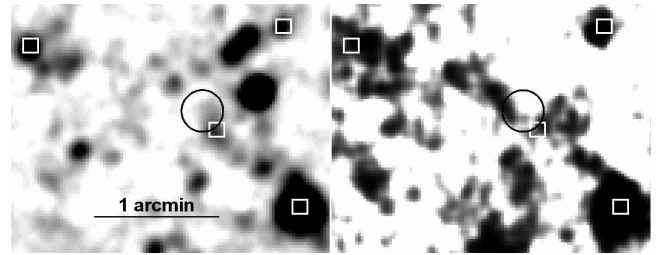}} \\
\caption{{\em WISE} W1 ($3.6$\,$\mu$m) and W3 (12\,$\mu$m) maps around the position of RX\,J2143.0+0654. The same field of view is shown as in the {\em Herschel} PACS maps of Figure~\ref{j2143H}. The circle with a radius of $10\arcsec$ marks the INS position, the white squares mark bright sources common in the {\em Herschel} PACS, {\em WISE} W1 and W2, and $H$-bands. 
\label{j2143W2}}
\end{center}
\end{figure}

\begin{table*}
\begin{center}
\caption{Flux measurements and limits \label{herschelres}}
\begin{tabular}{lccc|rr}
\tableline\tableline
Object & F$^{4.5\mu m}$ &  F$^{\rm blue}$  & F$^{\rm red}$ & M$_{\rm dust}^{\rm 20K}$& M$_{\rm dust}^{\rm 100K}$\\
& [$\mu$Jy] & [mJy] & [mJy] & [M$_{\earth}$]& [M$_{\earth}$]\\
\tableline\tableline
RX\,J0420  & $<2.1$ & $<4.5$ & $<7.0$ & $<1.0$ & $<0.02$\\
RX\,J0720$^{\rm a}$  & $<3.5$ & $<5.2$ & $<9.4$ & $<1.4$ & $<0.02$ \\
RX\,J0806  & $<11.4^{\rm PD,CN}$ & $<4.9$ & $10 \pm 5$$^{\rm CN}$ & $0.72\pm 0.36$ & $0.012\pm 0.006$\\
RX\,J1308  & $<1.2$ & $<1.7$ & $<5.2$ & $<1.5$ & $<0.02$\\
RX\,J1605  & $<2.7$ & $<6.1$ & $<12.2$ & $<2.2$ & $<0.04$\\
RX\,J1856  & $<8.9^{\rm CN}$ & $<7.7$ & $<7.8$& $<0.2$ & $<0.004$  \\
PSR\,J1848 & $\cdots$ & $<3.6$ & $<5.9$ & $<6.2$ & $<0.1$\\
RX\,J2143  & $<2.3$ & $<5.0$ & $7.8 \pm 4.8$$^{\rm CN}$ & $1.7 \pm 1.0$  & $0.03 \pm 0.02$ \\
\tableline\tableline
\end{tabular}
\end{center}
\tablecomments{The results of the following bands are listed: {\em Spitzer} IRAC2 (4.5$\mu$m), {\em Herschel} PACS blue band (60--85\,$\mu$m), and red band (130--210\,$\mu$m). For {\em Spitzer} data, the  color corrections for a $T=200$\,K blackbody are applied, but note the general remarks regarding color corrections in Section~\ref{datared}. The superscript $^{\rm PD}$ indicates a potential (weak) detection, $^{\rm CN}$  indicates  confusing neighbor sources. Dust masses are calculated using Equation~(\ref{dustmass}) for dust temperatures of 20\,K and 100\,K, the {\em Herschel} PACS red band measurements, and the distances as listed by \citet{Kaplan2009a}. All limits and errors are $5\sigma$ values. \protect\\
$^{\rm a}$ RX\,J0720 also has a {\em Spitzer} IRAC1 (3.6$\mu$m) limit, $F^{3.6\mu m} < 3.2$\,$\mu$Jy (color corrected for $T=200K$).}
\end{table*}

The field around RX\,J2143.0+0654 is densely populated in the infrared and includes several common sources in the {\em Herschel} PACS bands, {\em Spitzer}  IRAC, the {\em WISE} W1 \& W2 bands, and in the VLT ISAAC $H$-band observations of \citet{Posselt2009}; see Figures~\ref{j2143s} to \ref{j2143W2}. This allows us to confirm that the absolute pointing error of the {\em Herschel} observation is not larger than the expected value of $1\farcs{1}$.\\

There is emission in the red PACS band at the position of the INS, consisting of a blend of at least one southern and one brighter northern source (see Figure~\ref{j2143H}, the southern source is marked with a box and was also detected by {\em WISE}). In the blue PACS band, there is very faint emission northwest of the INS.  
The peak position of the $160$\,$\mu$m emission is $3\farcs{7}$ north of the optical INS position, about $2\farcs{6}$ from the closest N(IR) source.
Considering the unknown NS proper motion, we cannot ruled out the possibility that this source is the counterpart of the INS. However, it appears unlikely that RX\,J2143.0+0654 has the implied large proper motion ($>0\farcs{6}$\,yr$^{-1}$). While the INS distance is unknown, it is expected to be larger than 300\,pc \citep{Posselt2007}. For comparison, the closest of the XTINSs, RX\,J1856.5--3754 ($D=140$\,pc) has also the largest proper motion (330\,mas\,yr$^{-1}$) of those four of the XTINSs for which proper motion measurements or limits were reported.\\

A part of the red band emission at the INS position is likely to come from the bright infrared southwestern source which is seen in several bands. Given the proximity of the sources, their faintness and the FWHM in the red band of $10\farcs{5} \times 12\farcs{02}$, it is not possible to remove the sources with the necessary accuracy to search for residual $160$\,$\mu$m emission at the optical position of the INS. 
We used an aperture with radius of $5\arcsec$ at the position of INS to measure the flux at this position. The aperture-corrected flux is $F^{\rm red}= 7.8 \pm 4.8 \, (5\sigma)$\,mJy. 
Due to the surrounding sources, this value must be regarded with caution and it can be an overestimate of any actual flux from the INS position.  
In Section~\ref{alternatives} we discuss the likelihood of the {\em Herschel} PACS red band emission being not associated with the NS.\\

In the blue band  there is no prominent source emission at the INS position. Using apertures in different source-free regions and at the target position, we determined the $5\sigma$ upper limit for the RX\,J2143.0+0654 flux as $F^{\rm blue}_{5\sigma} = 5.0$\,mJy .\\

We used the VLT ISAAC $H$-band image by \citet{Posselt2009}, {\em Spitzer}, and 2MASS point sources for cross-checks with the {\em WISE} W1-W3 data, see Figures~\ref{j2143s} and \ref{j2143W2}. 
There is no noticeable astrometric shift between these bands. 
The W1 emission west of the INS position is probably related to the southwestern sources seen with {\em Spitzer}, and no enhanced emission is detected at the INS position. 
In W3, there is no  emission peak at the INS position, although there is an eastern source at $\approx 8\arcsec$ and a faint southern source at $\approx 5\arcsec$ from the nominal NS position.  
Since the astrometry of sources in these {\em WISE} images agree well with those of N(IR) sources, we conclude that RX\,J2143.0+0654 is undetected in the {\em WISE} W1-W3 data. 

\section{Discussion}
\subsection{The likelihood for the {\em Herschel} emission to be associated with field sources}
\label{alternatives}
In Sections~\ref{sub0806} and \ref{sub2143}, we reported on {\em Herschel} red band emission at or near the positions of RX\,J0806.4--4123 and RX\,J2143.0+0654.
How likely is the emission in the {\em Herschel} PACS red band associated with a source different from the INS ?
Confusion from faint Galactic stars is negligible for the PACS red band\footnote{herschel.esac.esa.int/Docs/HCNE/pdf/HCNE\_ScienceDoc.pdf}. The main two components of sky background confusion noise in the long-wavelength bands of {\em Herschel} are extragalactic sources and interstellar cirri.
\citet{Sibthorpe2013} showed that, for a given {\em Herschel} PACS flux level, results from typical extragalactic fields can be used to estimate the background source numbers in typical debris disk surveys. 
They used the results from the PACS Extragalactic Probe \citep{Berta2011, Lutz2011} to estimate the expected background source number density for given flux and flux uncertainty levels.
Following the approach by \citet{Sibthorpe2013}, we estimate the chance probability of detecting an $F_{\rm red} \approx 10$\,mJy extragalactic source in a circle with radius of $r=5\arcsec$ as $0.5$\,\%. 
For RX\,J2143.0+0654, we estimate the respective chance probability of detecting an $F_{\rm red} = 7.8 $\,mJy extragalactic source in a circle with radius of $r=5\arcsec$ as $0.7$\,\%. 
Given these chance probabilities, the binomial probability that we detect background sources in two of overall eight XTINS fields is at maximum 1\,\%, for $7.8$\,mJy sources. 
But the probability to detect one background galaxy at one of the eight XTINS positions is already 5\,\%.\\  

\begin{figure}[t]
\centering
\includegraphics[height=70mm,bb=36 16 570 540]{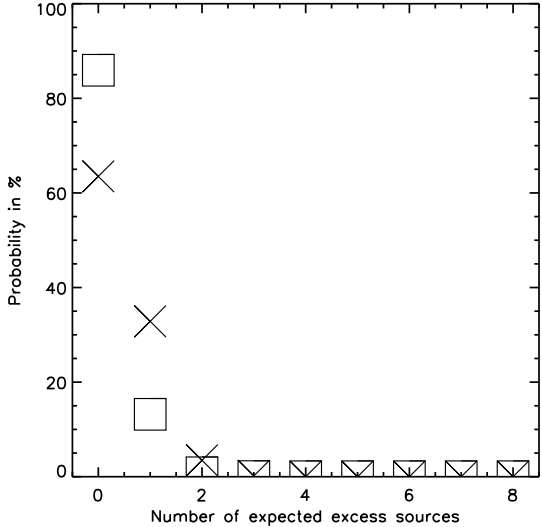}
\caption{The probability of detecting $N_{\rm excess}$  excess sources in our investigated NS sample. We use the Monte Carlo code by \citet{Gaspar2013}, simulating $10^6$ datasets of eight sources with our detection requirement $\gtrsim 5\sigma$ in a $r=5\arcsec$ photometry aperture. Boxes indicate the values for using the median interstellar cirrus noise level of our eight NS fields. Crosses indicate the results of the simulation which takes into account the exceptionally high ISM background flux for RX\,J0806.4--4123.
\label{ProbExcess}}
\end{figure}

\citet{Gaspar2013} argued that \citet{Sibthorpe2013} underestimate the confusion noise probabilities, mainly because galaxies fainter than a considered flux limit may also contribute to the overall confusion noise. 
We used the Monte Carlo Code by \citet{Gaspar2013} to estimate the probabilities for $N_{\rm excess}$ unrelated `excess sources' in the photometry apertures of our eight NSs. 
We realized $10^6$ datasets of eight sources with our detection requirement $\gtrsim 5\sigma$ (Table~\ref{herschelres}, i.e., 5\,mJy for RX\,J0806.4--4123) in a $r=5\arcsec$ photometry aperture.
In a nutshell, the code by \citet{Gaspar2013} simulates background sources in a single large field ($0.5$ square degrees), then properties for 8 random positions from this field are estimated 10$^6$ times. The latter step includes the consideration of noise from interstellar cirri by using a probability distribution with standard deviation $\sigma_{\rm cirrus}$ which is calculated from the observed far-infrared ISM flux background. For details on the code and the assessment of its statistical performance, we refer to \citet{Gaspar2013}.
For our simulations, we used a cirrus noise value of $\sigma_{\rm cirrus}=0.84$\,mJy, which corresponds to the median value of the ISM background flux of $11.74$\,MJy\,sr$^{-1}$, as estimated by using Hspot\footnote{The {\em Herschel} tools are available at: herschel.esac.esa.int/Tools.shtml} for the observing dates and positions of our eight NSs. 
The resulting probability plot is shown in Figure~\ref{ProbExcess}. The probability to detect no, one or two excess source(s) among our eight NSs is 86\,\%, 13\,\% and 1\,\%, respectively. Thus, it seems unlikely that both our detections are excess sources.\\

\citet{Kaplan2011} investigated RX\,J0806.4--4123 with the {\em Hubble Space Telescope} ({\em HST}) and found no source other than the NS  within $2\arcsec$ of the INS position using the Advanced Camera for Surveys, Wide Field Channel/filter F475W down to a ST magnitude of $27.92\pm 0.22$. Considering the {\em Herschel} pointing accuracy ($1\farcs1$, Section~\ref{sub0806}), its large FWHM ($\approx 11\arcsec$ for the red band), there are, however, several {\em HST} sources which currently cannot be excluded as potential {\em Herschel} background galaxy counterparts. 

There is also another alternative explanation for the {\em Herschel} PACS red band source:  interstellar cirrus. In fact, the ISM background flux in the direction of RX\,J0806.4--4123 is estimated by Hspot to be about a factor $10$ higher than the median value of the remaining seven NS fields. Illustrating the higher ISM background, the recent 3D maps of the local ISM distribution by \citet{Lallement2014} show a local ISM cloud clump in the direction of RX\,J0806.4--4123. A typical interstellar cirrus has been found to have temperatures of about $\sim 20$\,K with a range of about 4\,K around this value (\citealt{Veneziani2013}, see also \citealt{Gaspar2013}) -- a temperature range very reminiscent of the one inferred for our {\em Herschel} detection around RX\,J0806.4--4123 (see section~\ref{discussion1}).
If we consider explicitly the higher ISM background flux for this one source in the Monte Carlo simulation by \citet{Gaspar2013}, we obtain  63\,\%, 33\,\% and 3\,\% as the probabilities for detecting no, one, or two excess source(s) among our eight NSs, respectively. 
The probability of one excess source among the eight NSs has nearly tripled, but the probability for no excess source in the sample is still a factor 2 higher.\\

In the case of RX\,J2143.0+0654, the {\em HST} observations are not constraining with respect to excluding obviously present galaxies as counterparts to the faint {\em Herschel} emission either.
Regarding interstellar cirri, the ISM background flux in the direction of RX\,J2143.0+0654 is nearly the same as the median value for the seven NS fields (excluding RX\,J0806.4--4123). Thus, the probability for an excess source due to an ISM cirrus is less than in the case of RX\,J0806.4--4123.

\subsection{Multi-wavelength constraints on the dust emission}
\label{discussion1} 
In the following, we assume that the detected {\em Herschel} emission originates from dust associated with the XTINS.
We summarize the {\em Herschel} and {\em Spitzer}  results of the previous subsections in Table~\ref{herschelres}. 
Assuming the emitting dust to be optically thin at submillimeter wavelengths, we can use the {\em Herschel} 160\,$\mu$m measurements to calculate the dust mass assuming a single temperature for all dust grains:
\begin{equation}
M_d=\frac{F_{\nu} D^2}{B_{\nu}(T_d) {\kappa}^d_{\nu}},
\label{dustmass}
\end{equation}
where $\nu=1.9 \times 10^{12}$\,Hz is the frequency, $F_{\nu}$ is the measured flux density (limit), $D$ is the
distance, $B_{\nu}(T_d)$ is the Planck function at a dust temperature $T_d$, 
and $ {\kappa}^d_{\nu}$ is the dust mass absorption coefficient. 
We use $ {\kappa}^d_{\rm 160\mu m}=13$\,cm$^2$\,g$^{-1}$ 
for the PACS red band, which we calculated from the optical constants by \citet{Dorschner1995} assuming dust grains consisting of amorphous magnesium silicate with a grain density $\rho_g = 2.7$\,g\,cm$^{-3}$. 
The dust composition and, therefore, the dust mass absorption coefficient is in general highly uncertain (by a factor 3 to 5, e.g., \citealt{Posselt2010}).
Additionally, the dust likely has a temperature distribution, not a single temperature. The composition and dimensions of a fallback disk around a $\sim 1$\,Myr old NS are unknown, too. Thus, the derived  dust masses in Table~\ref{herschelres} and Figure~\ref{summaryrxj0806} are only crude estimates.\\

\begin{figure*}
\noindent\begin{minipage}[b]{.33\textwidth}
\begin{center}
\hspace{1cm}RX\,J0420.0--5022
\includegraphics[width=64mm]{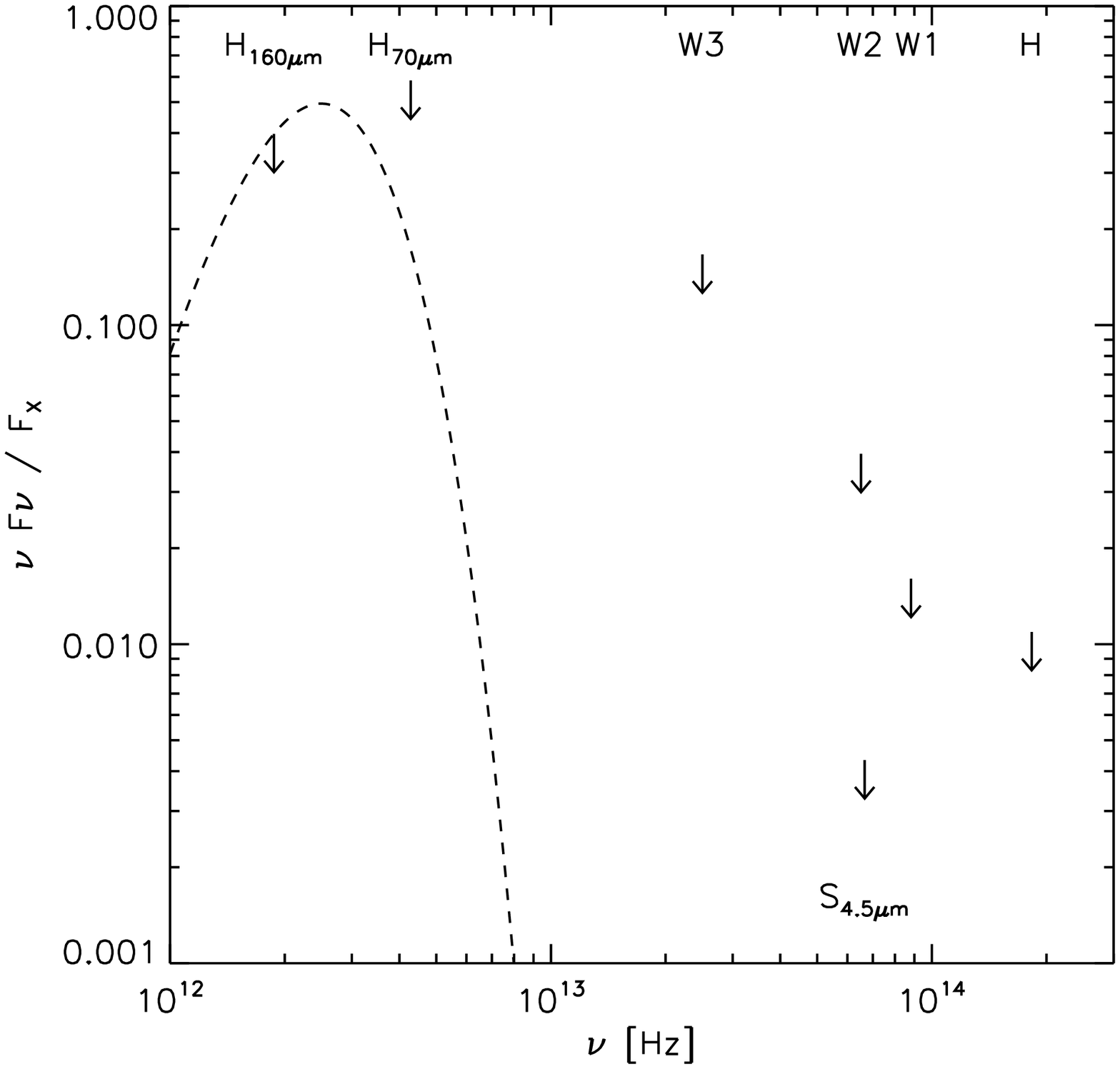}
\end{center}
\end{minipage} 
\begin{minipage}[b]{.33\textwidth}
\begin{center}
\hspace{1cm}RX\,J0720.4--3125
\includegraphics[width=64mm]{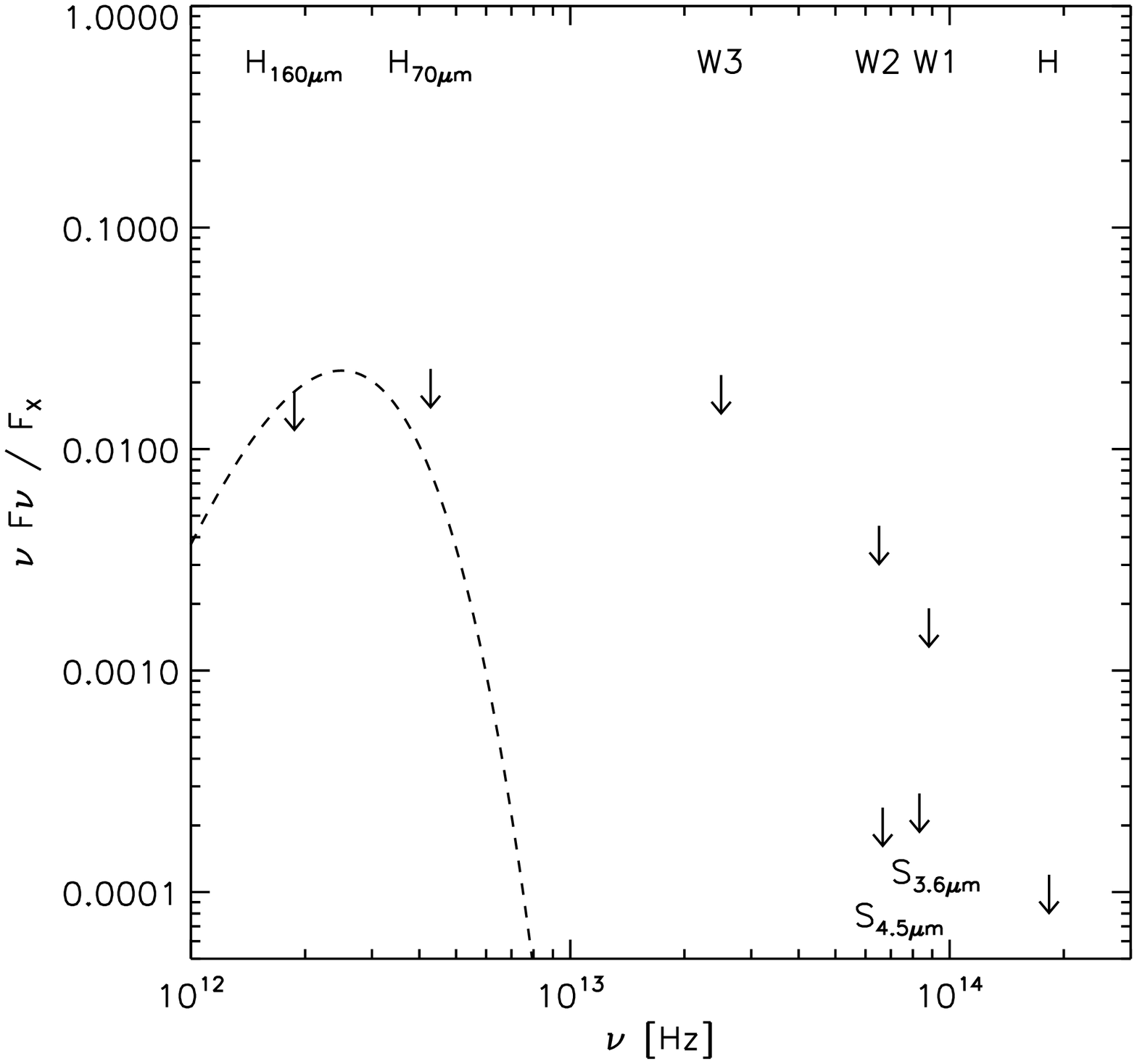}
\end{center}
\end{minipage}
\begin{minipage}[b]{.33\textwidth}
\begin{center}
\hspace{1cm}RX\,J1308.6+2127
\includegraphics[width=64mm]{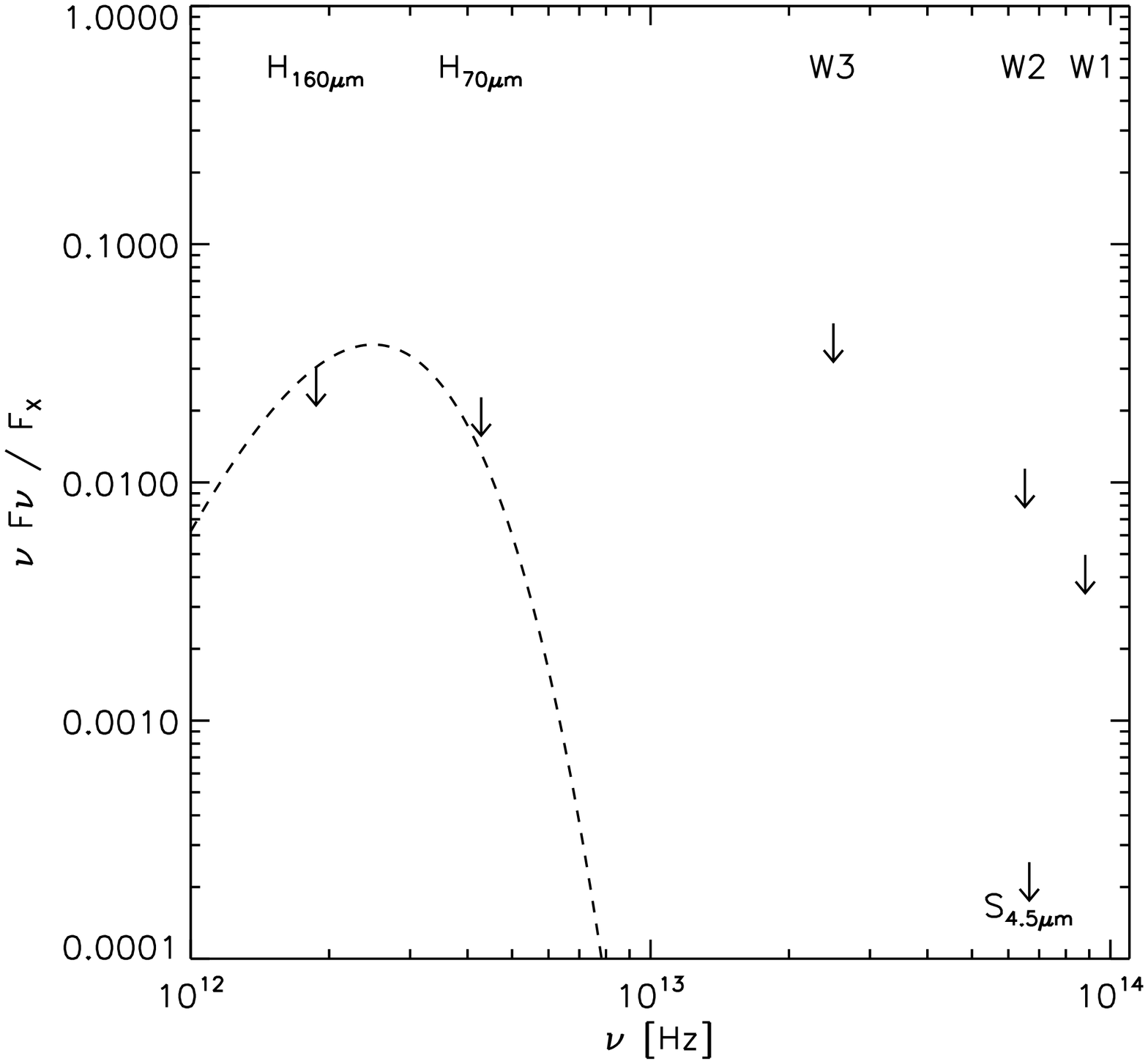}
\end{center}
\end{minipage}\\ 
\noindent\begin{minipage}[b]{.33\textwidth}
\begin{center}
\hspace{1cm}RX\,J1605.3+3249
\includegraphics[width=64mm]{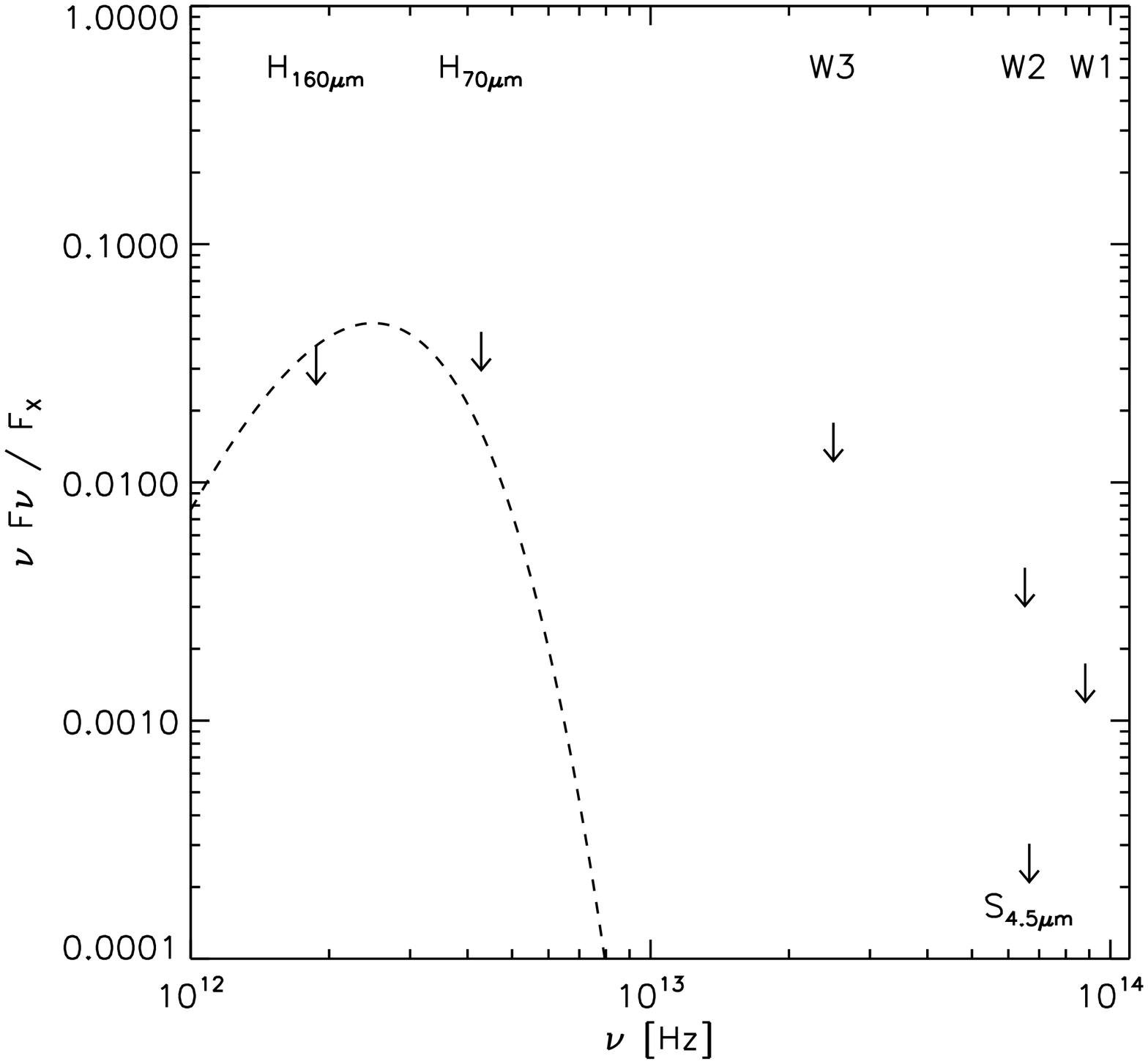}
\end{center}
\end{minipage}
\begin{minipage}[b]{.33\textwidth}
\begin{center}
\hspace{1cm}RX\,J1856.5--3754
\includegraphics[width=64mm]{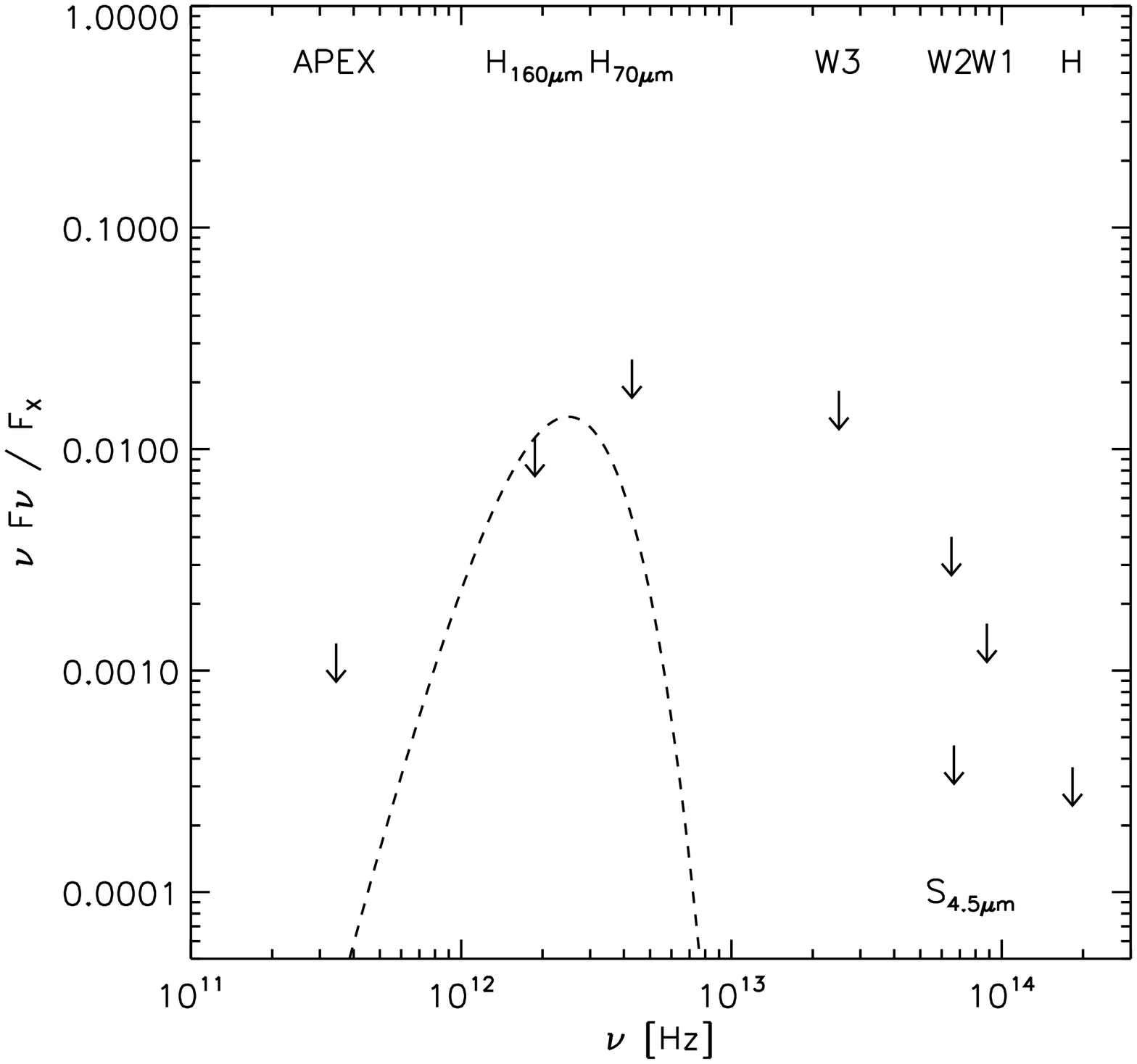}
\end{center}
\end{minipage}
\begin{minipage}[b]{.33\textwidth}
\begin{center}
\hspace{1cm}RX\,J2143.0+0654 
\includegraphics[width=64mm]{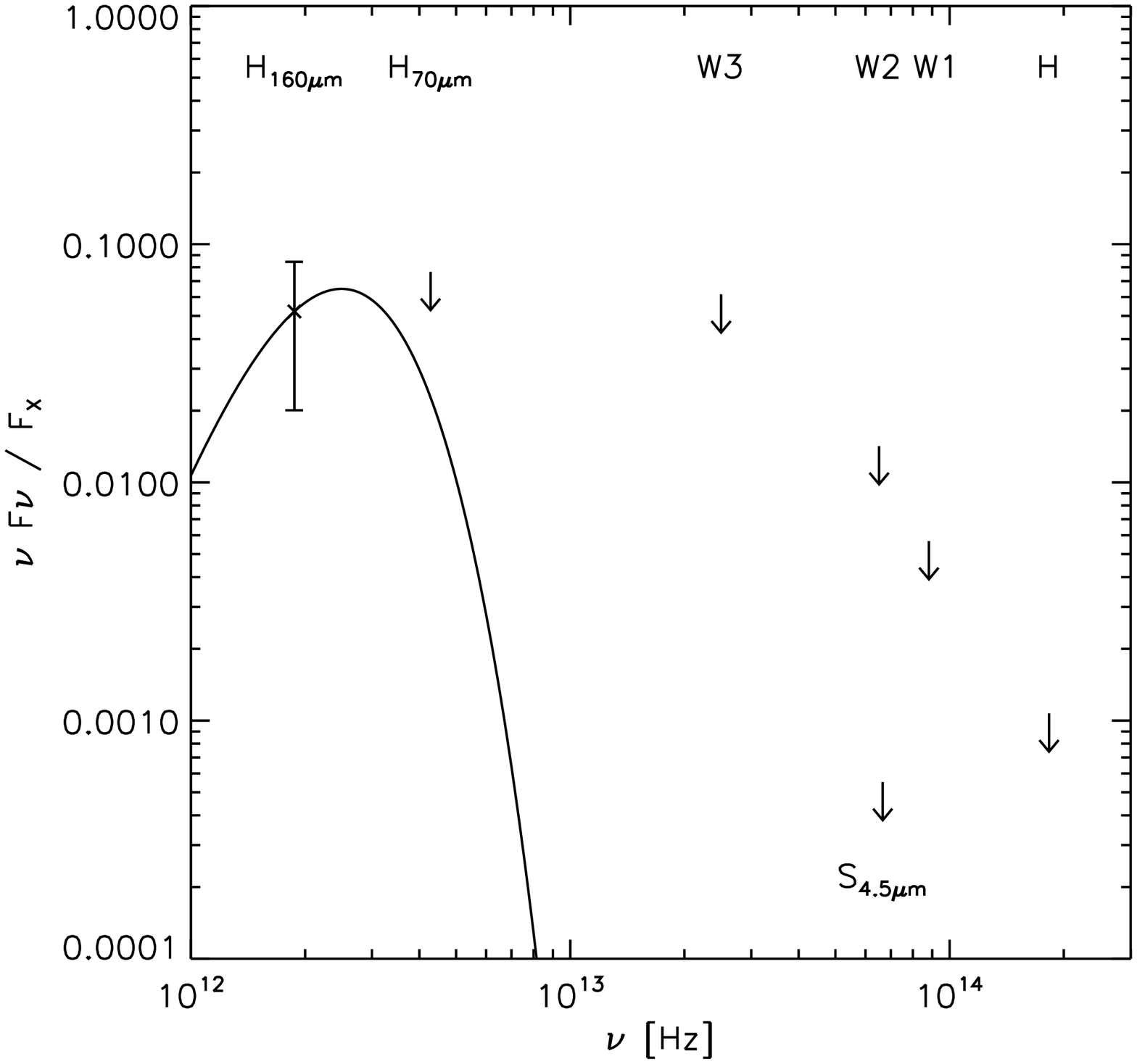}
\end{center}
\end{minipage} 
\hfill
\caption{Measured values and upper limits for the IR/submillimeter to X-ray flux ratios for six XTINSs. The X-ray fluxes are for the energy range 0.1--2.4\,keV and are taken from \citet{Haberl2013}. Note that the XTINSs show only very soft thermal X-ray emission. The infrared/submillimeter limits/values are taken from  Tables~\ref{WISElimits} and \ref{herschelres} and the measurements by \citet{Posselt2010,Posselt2009}. 
For {\em Spitzer} and {\em WISE} data, exemplary color corrections for a $T=200$\,K blackbody are applied, but note the general remarks regarding color corrections in Section~\ref{datared}.
No extinction correction has been applied since the XTINSs have very small absorbing hydrogen column densities ($N_H \approx 10^{20}$\,cm$^{-2}$), corresponding to negligible extinction corrections in the Infrared ($A_H\approx 0.01$, $A_{\rm 3.6\mu m} \approx 0.004$; following the extinction relations by \citealt{Vuong2003} and \citealt{Indebetouw2005}). The dashed lines show upper limits on the emission of dust with a temperature $T_d=20$\,K according to Equation~(\ref{Fnulonglambda}). The solid line in the case of RX\,J2143.0+0654 indicates the respective constraints from the measured {\em Herschel} aperture flux.}
\label{fluxplots}
\end{figure*}

In Figures~\ref{fluxplots} and \ref{summaryrxj0806}, we show the ratios of the IR/submillimeter fluxes (limits) to the X-ray fluxes of the XTINSs. The obtained flux ratios (limits) are usually in the range $10^{-4}$ to $10^{-3}$ for the $H$-band and the {\em Spitzer} $4.5$\,$\mu$m data, but $0.01$ to $0.1$ for the {\em Herschel} data.
For comparison, we refer to similar plots for three CCOs by \citet{Wang2007}  (for optical and IR fluxes only). At the {\em Spitzer} wavelengths, flux ratio limits of $10^{-2}$ to $10^{-4}$ were reached for the CCOs, but no disks were detected. \citet{Wang2007} also showed a comparison plot for the only magnetar where a disk is thought to be detected -- 4U\,0142+61 \citep{Wang2006}. For this magnetar, the flux ratio at {\em Spitzer} IRAC wavelengths is $\approx 6 \times 10^{-5}$. This comparison might indicate that our limits are simply not deep enough. On the other hand, the single object 4U\,0142+61 might not be defining for all potential NS disks. 
Our {\em Herschel} detections of emission around RX\,J0806.4--4123 and RX\,J2143.0+0654 could indicate predominantly cold dust around these (older) NSs. {\em Spitzer}  $4.5$\,$\mu$m observations are not sensitive enough to detect such cold dust.\\

\begin{figure*}[t]
\noindent\begin{minipage}[t]{.33\textwidth}
\begin{center}
\includegraphics[width=60mm]{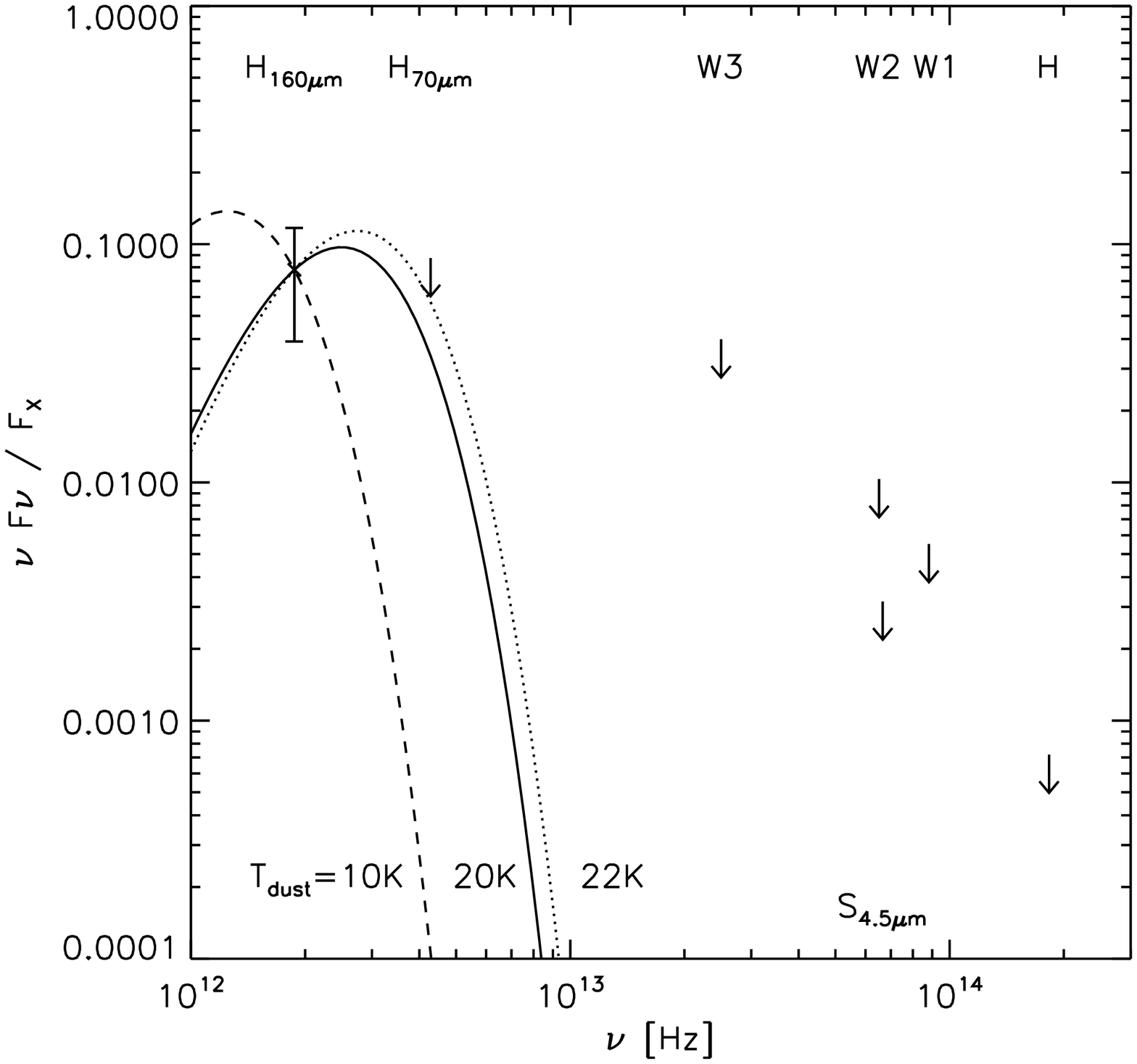}
\end{center}
\end{minipage}
\hspace{-0.5cm}
\begin{minipage}[t]{.33\textwidth}
\begin{center}
\includegraphics[width=60mm]{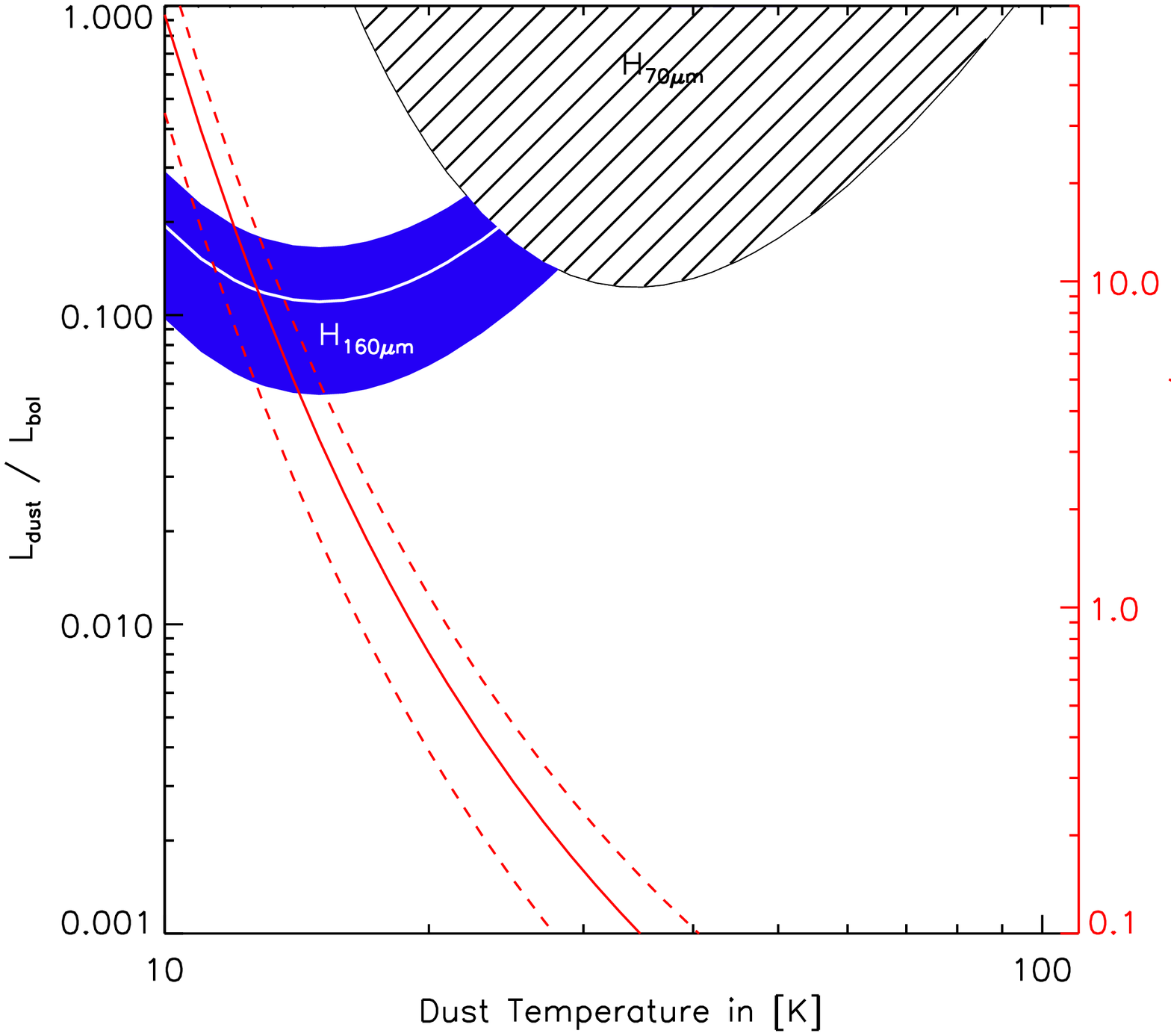}
\end{center}
\end{minipage}
\hspace{0.2cm}
\begin{minipage}[t]{.33\textwidth}
\begin{center}
\includegraphics[width=60mm]{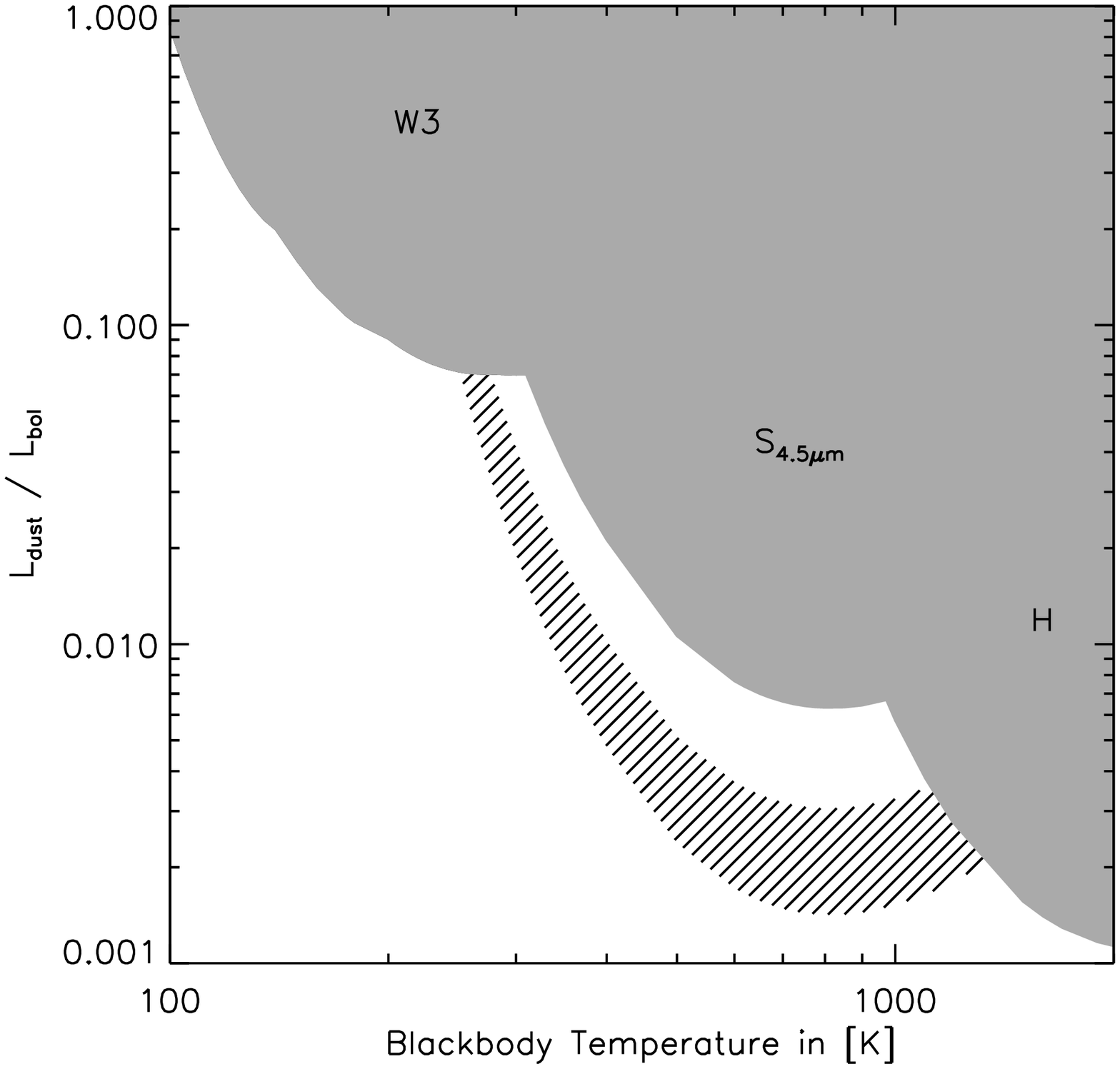}
\end{center}
\end{minipage}
\hfill
\caption{Measured values and upper limits for the IR/submillimeter to X-ray flux ratios for RX\,J0806.4--4123 (left panel) and constraints on the luminosity and temperature of dust around it (middle and right panel). The flux ratios are from this paper and the literature as described for Figure~\ref{fluxplots}. The luminosity ratios are derived with respect to the bolometric luminosity (which is basically the X-ray luminosity), the mass is calculated using a distance of 250\,pc which is based on the X-ray absorption \citep{Kaplan2009a, Posselt2007}.
The middle panel shows the luminosity constraints for dust emission according to Equation~(\ref{lumfnurel}). The blue area marks the luminosity ratio range implied by the {\em Herschel} $160$\,$\mu$m measurement (white line) and its $5\sigma$ uncertainty, the dashed area indicates the excluded dust luminosities according to our {\em Herschel} blue band observations. The red solid and dashed lines show the inferred mass constraints from the  {\em Herschel} $160$\,$\mu$m measurement. The right panel shows the upper limit on the luminosity ratios as implied for blackbody emitters under optically thin conditions. The shaded area indicates the excluded region for blackbody emitters. Similarly, the hatched area shows  the allowed range for the marginal $1.9\sigma$ {\em Spitzer} $4.5$\,$\mu$m detection, see Section~\ref{sub0806}. Realistic dust disk models require the application of the Mie theory and would result in lower limits in the right panel (see text).\vspace{0.5cm}}
\label{summaryrxj0806}
\end{figure*}

Dust emission at long wavelengths can be described in the Rayleigh limit if $\lambda > \max{( 2  \pi a, \lambda_{\mathrm{Res}}})$, where $a$ is the (spherical) grain size radius, and $\lambda_{\mathrm{Res}}$ is the largest resonance wavelength of the complex refractive index as tabulated in the DOCCD\footnote{Database of Optical Constants for Cosmic Dust  www.astro.uni-jena.de$/$Laboratory$/$OCDB$/$index.html}.
If this condition is fulfilled, the flux density $F_{\nu} \propto \nu^2$ (see Appendix~\ref{dusttemp}), and assuming optically thin conditions for the grain as well as the overall medium (e.g., dusty disk or cloud), we can write:
\begin{equation}
F_{\nu}=\frac{B_{\nu} (T_d)}{B_{\nu_0} (T_d)} \left(\frac{\nu}{\nu_0}\right)^2 F_{\nu_0},
\label{Fnulonglambda}
\end{equation}
where $\nu_0 \ll 10^{13}$\,Hz is the reference frequency. Here, we choose the {\em Herschel} red band ($\nu_0=1874$\,GHz) as reference band and show the constraints on the expected dust fluxes at long wavelengths in Figures~\ref{fluxplots} and \ref{summaryrxj0806}. In the former we use an exemplary temperature $T_d=20$\,K, in the latter we illustrate the effect of choosing different temperatures.\\

In the wavelength regime of W1-W3, {\em Spitzer} IRAC and the $H$-band, the dust absorption and emission properties are highly non-monotonic functions of $\lambda$ (see, e.g, the laboratory absorption coefficient measurements of dust grains by \citealt{Dorschner1995}), and the dust medium is in general not optically thin for (its own) emission at such wavelengths. Realistic flux models would require the application of the Mie theory for the radiation transport calculation for a distribution of dust grains. Given that we deal only with flux limits at these wavelengths, such modeling is beyond the scope of this paper. To estimate at least the maximal upper limits on the luminosity ratios, we assume blackbody emission.  
For blackbody emission, the dust luminosity limits, $L_{\rm dust}(T_d)$, are 
related to the measured flux density limits, $F_{\nu}$, 
\begin{equation}
F_{\nu} =\frac{B_{\nu}(T_d) L_{\rm dust} }{4 D^2 \sigma _{SB} {T_d}^4 },
\end{equation} 
where $\sigma _{SB}$ is the Stefan-Boltzmann constant. 
The right plot in Figure~\ref{summaryrxj0806} shows the blackbody luminosity constraints for RX\,J0806.4--4123.\\

For $h\nu_0 \gg 2.8 kT$, optically thin dust conditions, and uniform properties of the dust grains as well as an uniform temperature, we derive for the total luminosity of the dust:
\begin{eqnarray}
L_{\rm dust}(T_d)=
\frac{160}{21}\pi^2 \left(\frac{kT_d}{h\nu_0}\right)^2 \kappa ^d_{\nu_0} M_d \sigma _{SB} T_d^4 \nonumber \\ =
F_{\nu} D^2 \frac{160}{21} \pi^2 \left(\frac{k}{h\nu}\right)^2 \frac{\sigma _{SB}}{B_{\nu} (T_d)} T_d^6,
\label{lumfnurel}
\end{eqnarray}
where $k$ is the Boltzmann constant, and $h$ is the Planck constant. The middle plot of Figure~\ref{summaryrxj0806} shows the constraints on $L_{\rm dust}(T_d)/L^{NS}_X $ from the {\em Herschel} red band value and blue band limit. We see that only low dust temperatures ($T_d<25$\,K) are allowed considering the detection and non-detection in the red and blue band, respectively. \\

\subsection{Constraints on a potential dusty disk or torus}
\label{discussion2}
In the previous Section we discussed the constraints on the dust associated with the XTINSs only in general.
Here, we discuss constraints on different properties  of potential disks assuming that the detected {\em Herschel} fluxes are indeed due to emission from dust {\emph{around}} the NSs.
\citet{Bryden2006} noted that only the brightest debris disks detected with {\em Spitzer} have $L_{\rm dust}/L_{\rm bol,\ast} \geq 10^{-4}$. The recent {\em Herschel} DUNES survey by \citet{Eiroa2013} reported  $L_{\rm dust}/L_{\rm bol,\ast}$ ranging from $10^{-6}$ to $3\times 10^{-4}$ for {\em Herschel} debris disk candidates around solar-type stars.
If the {\em Herschel} $160$\,$\mu$m detections for two 2 XTINSs and the marginal {\em Spitzer} $4.5$\,$\mu$m and $H$-band detections of emission around RX\,J0806.4--4123 are confirmed to come from disks, the luminosity ratios would be two to three orders of magnitudes higher than the values seen for non-degenerate stars. 
The implied large absorption cross section of the disk appears questionable. 
In contrast to the luminosities of main sequence stars, most of the XTINS's luminosity is emitted in soft X-rays ($<1$\,keV). In contrast to other wavelengths, soft X-ray photons are very efficiently absorbed and re-emitted by all dust grains ($\approx 100$\,\% of each incident soft X-ray photon; e.g., \citealt{Dwek1996}). 
The absorption cross section of a putative disk cannot, however, be larger than the geometric cross section of that disk.
Hence, even if one accounts for an exceptionally high dust heating efficiency, geometric arguments exclude a thin flared disk from gobbling up 20\% of the stellar luminosity. The implied geometric shape for an assembly of dust grains surrounding the NS would rather be a dusty torus or an envelope.\\  

We use the Rayleigh approximation to estimate the possible location of the dust around the NS.
As outlined in the Appendix~\ref{dusttemp}, the temperature distribution, $T_\mathrm{g}(r)$, of  amorphous magnesium silicate dust grains with sizes $a$ can be calculated as: 
\begin{equation}
T_\mathrm{g}(r)= 39 \left(\frac{L_{\mathrm{bol},30}}{a_{\mu \mathrm{m}}} \right)^{1/6} \left(\frac{1}{r_{14}} \right)^{1/3}\,\mathrm{K},
\label{tg}
\end{equation}
where the NS luminosity, $L_{\mathrm{bol},30}$, is normalized to $10^{30}$~erg~s$^{-1}$, and the distance, $r_{14}$, between the NS and the grain is normalized to  $10^{14}$\,cm. Note that this equation is only valid for $T<\min{(96\,\mathrm{K},2300 a^{-1}_{\mu \mathrm{m}}\,\mathrm{K})}$.

As shown in Section~\ref{discussion1} and Figure~\ref{summaryrxj0806}, the inferred temperature for potential circumstellar dust around, e.g., RX\,J0806.4--4123 is in the range of 10 to 22\,K.  A dust grain with such temperatures would be expected at radii of $r_d= 2.3 \times 10^{16} a_{\mu \mathrm{m}}^{-1/2}$\,cm to $ 2.2 \times 10^{15} a_{\mu \mathrm{m}}^{-1/2}$\,cm, corresponding to $1600\, a_{\mu \mathrm{m}}^{-1/2}$\, A.U. and $150 \, a_{\mu \mathrm{m}}^{-1/2}$\,A.U., respectively.  
At a distance of $\approx 250$\,pc, this translates into angular sizes of about $6\arcsec$ and $1\arcsec$, i.e., the source would be still unresolved in the {\em Herschel} red band which is consistent with the observations. 
Thus, we cannot rule out such temperatures based on the expected emission extension for RX\,J0806.4--4123.
For RX\,J2143.0+0654, the inferred radii would be factor 2.8 larger due to its higher X-ray luminosity, although the angular scales are probably similar because of the likely larger distance of this XTINS.
The implied orbital radii of the $22$\,K dust grains with sizes $a\lesssim 1$\,$\mu$m, are larger than the radii commonly considered for (young and gas-rich) fallback disks, $R_{\rm FD} \approx 10^{10} - 10^{14}$\,cm (e.g., \citealt{Perna2014}). 
After removal of the gas in the possible disks around the relatively old INSs, however, the dust is  difficult to remove by ISM drag (see also below), and a pure dust or debris disk could, in principle, expand to larger radii.
Looking at our own solar system, there are the following reminiscent large structures: the Kuiper belt at $30-60$\,A.U. with lateral coverage of up to $40^{\circ}$ (e.g., \citealt{Brown2001}) and the hypothesized near-spherical Oort cloud with radii from  $10^4 - 10^5$\,A.U. (e.g., \citealt{Dones2004}). The formation of neither the Kuiper belt nor the Oort cloud are fully understood, but in general it is believed that gravitational interactions of belt/cloud objects with the newly-formed giant planets were of major importance \citep{Morbidelli2008,Dones2004}. Survival or formation of giant planets around a NS are unlikely, and a mechanism for the extension of an initial compact fallback disk would be needed if the emitting dust grains are smaller than $10$\,$\mu$m. Constraints on the emitting grain size could be obtained by additional investigations at submillimeter wavelengths. \\

\begin{figure}
\includegraphics[width=80mm]{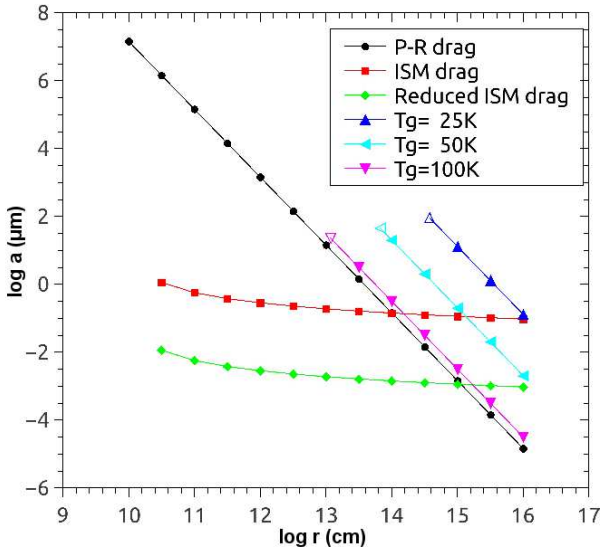}
\caption{Distribution of spherical grains with radius $a$ in $\mu$m over the distance from the NS, $r$ in cm, for different temperatures and dust removal effects.
The blue, cyan, and magenta (all with triangles) lines indicate the temperature a dust grain would
have if Eq.~\ref{tg} is applied. The open triangle symbols indicate the region left of which the conditions for Eq.~\ref{tg} are not met.
The black line indicates removal of grains due to the Poynting-Robertson
drag.
Grains {\emph{below}}
the black line are expected to be removed at the assumed disk age of 0.5 Myr.
Similarly, the green and red lines indicate the effect of the ISM drag
the dust grains experience when they move together with the NS disk through
the ISM (e.g., \citealt{Phillips1994}), please see also text.
\label{prdrag}}
\end{figure}

The smallest grains are the hottest and contribute most to the observed dust luminosity.
If, in the case of cicumstellar dust, already $\approx 20$\,\% of the NS luminosity is re-emitted by cold dust, but the upper limit at {\em Spitzer} wavelength is less than  1\,\% (because of blackbody assumption; see Figure~\ref{summaryrxj0806}), it follows that there are no significant amounts of small hot dust grains around RX\,J0806.4--4123. 
An explanation for this finding can be the effect of the Poynting-Robertson (P-R) drag (e.g., \citealt{Phillips1994}). 
In Figure~\ref{prdrag}, we plot lines corresponding to the P-R drag and different dust
temperatures (for the latter we use Eq.~\ref{tg}). The black line corresponds to the P-R drag limit for a putative disk around a NS with radius of $R_{\rm NS}=10$\,km and a disk age of 0.5\,Myr. Dust grains below that line are removed for the respective radius in a putative disk. 
We see that small grains close to NS are effectively eliminated.
Dust grains with size $\sim 1 \mu$m can exist only at radii $r > 3 \times 10^{13}$~cm.
In Figure~\ref{prdrag}, we also plot the effects of the ISM drag experienced by the dust grains when they move together with the NS disk/belt through the ISM (e.g., \citealt{Phillips1994}). The ISM drag depends on the inclination of the putative disk/belt with respect to the direction of proper motion. We use different inclination angles. If the putative disk moves edge-on in the direction of the proper motion, the inclination angle is  $0^\circ$. This assumption allows us to obtain an upper limit for the smallest grain size, represented by the red line in Figure~\ref{prdrag}. 
An inclination angle of $89^\circ$ corresponds to the putative disk moving nearly face-on in the direction of the proper motion. This scenario is more likely because many NSs have a spin-velocity alignment and angles $<10^\circ$ are most probable. In Figure~\ref{prdrag}, this scenario is represented by the green line. We assumed a disk age of 0.5\,Myr, a NS velocity of 100\,km\,s$^{-1}$, a number
density of ISM gas particles $1$\,cm$^{-3}$, and a dust grain density of 3\,g\,cm$^{-3}$ for our calculations.
The ISM drag is the dominating force on the dust grains at large distances. Similarly to the P-R line, grains below the ISM drag lines in Figure~\ref{prdrag} would be removed.
In the case of face-on disk motion, this mechanism would remove grains smaller than $\sim 1 \mu$m at all distances.
Overall, it appears likely that any dust around RX\,J0806.4--4123 consists of predominantly large, $a>1$\,$\mu$m, grains -- if the {\em Herschel} emission indeed comes from a dusty torus around the XTINS.\\

\section{Conclusions}
Using {\em Herschel} PACS, we detected $160$\,$\mu$m emission close to positions of two out of the eight investigated NSs. {\em Herschel} PACS $70$\,$\mu$m, {\em WISE} and {\em Spitzer} IRAC observations resulted only in upper flux limits for the positions of the eight NSs.
The ratios of the $160$\,$\mu$m band luminosity to the bolometric luminosity of the respective XTINSs are between 5\,\% and 20\,\%. If these detections are associated with the NSs, they would imply cold ($T_d \approx 20$\,K) dusty tori/belts around the NSs. 
For RX\,J0806.4--4123, the implied belt radius would be within the range discussed for fallback disks if only large ($a>10$\,$\mu$m) dust grains are present.
The implied dust torus radius is larger for smaller grains (e.g., $r \approx 10^{15}$\,cm for $a=1$\,$\mu$m), raising questions about the formation mechanism of such a dust belt. The higher X-ray luminosity of RX\,J2143.0+0654 would even imply a factor 3 larger torus radius.\\

There is a 3\,\% probability that both {\em Herschel} PACS red band detections are unrelated to the XTINSs, and there is a 33\,\% probability that one of the two is unrelated to the respective XTINS. 
The relatively large offset of the faint {\em Herschel} emission in the case of RX\,J2143.0+0654 and the possibility of a galaxy counterpart suggest that this detection may be associated with a background galaxy. 
For RX\,J0806.4--4123, an interstellar cirrus is an alternative to the dusty-torus hypothesis. Deeper observations at shorter and longer wavelengths with better spatial resolution could probe whether or not the {\em Herschel} $160$\,$\mu$m is associated with the NS.\\

For the six other NSs, we do not find any significant warm or cold dust emission. 
The reached flux ratio limits come close to those of bright debris disks around non-degenerate stars considering the warm dust, but for cold dust the {\em Herschel} observations are still not sensitive enough to exclude cold disks.
In principle, our observations are still consistent with the presence of cold dusty disks around the NSs if one relies on the flux ratios found for other debris disks which are at least two orders of magnitudes smaller than what we achieved here.    
There is, of course, the other possibility that the XTINS do not harbor disks, either because they never had or because they lost them due to the composition of the dust in the disk (e.g., only small dust grains close to the NS).
Currently, we cannot differentiate between the cold-disk and no-disk scenarios.
Future submillimeter interferometer observations provide an opportunity to test whether or not cold disks are present around these fascinating NSs.

\acknowledgments
We would like to thank  Herv\'{e} Bouy, Volker Tolls, and the staff from the NHSC Helpdesk for their helpful advice regarding {\em Herschel} data reduction. We would also like to thank Eliahu Dwek, Andr\'{a}s G\'{a}sp\'{a}r, Paola Popesso, and Bruce Sibthorpe for detailed explanations and code sharing, and the referee for valueable comments.

This work is based on observations made with {\em Herschel}, a European Space Agency Cornerstone Mission with significant participation by NASA. Support for this work was provided by NASA through award RSA 1437567 issued by JPL/Caltech.

This work is based in part on observations made with the {\em Spitzer Space Telescope}, which is operated by the Jet Propulsion Laboratory, California Institute of Technology under a contract with NASA. Support for this work was provided by NASA through award RSA 1416202 issued by JPL/Caltech.

S.B.P. was supported by the grant RFBR 12-02-00186.

This publication makes use of data products from the {\em Wide-field Infrared Survey Explorer}, which is a joint project of the University of California, Los Angeles, and the Jet Propulsion Laboratory/California Institute of Technology, funded by the National Aeronautics and Space Administration.

\bibliographystyle{apj}
\bibliography{Herschelbib}

\appendix
\section{A. Additional results on individual sources}
\label{detresults}
The following subsections present detailed results from the {\em WISE, Spitzer} and {\em Herschel} observations for the NSs without any significant IR detections. In contrast to Table~\ref{herschelres}, no color correction has been applied to the quoted flux limits. 
The Figures have different (manually tweaked) color scales and smoothing (only in the case of {\em Spitzer}; usually gaussian with a kernel radius of 3\,pixel), optimized to emphasize faint fluxes in the surrounding of the NS position. Circles in the pictures do not represent error circles. In the case of {\em Spitzer}, $r=2\farcs{4}$ (2 native IRAC pixel), the circle has the size of the used photometry aperture. In the case of {\em Herschel} and {\em WISE}, $r=10\arcsec$, they merely indicate the NS position. For an overview about the instrumental FWHM in each individual observing band, we refer to Sections~\ref{obs} and \ref{wise}.
Position uncertainties of the targets and instrumental pointing errors are discussed in Section~\ref{obs} as well as below if necessary.

\subsection{A.1. RX\,J0420.0--5022}
\label{sub0420}

\begin{figure}[h]
\begin{center}
{\includegraphics[width=85mm]{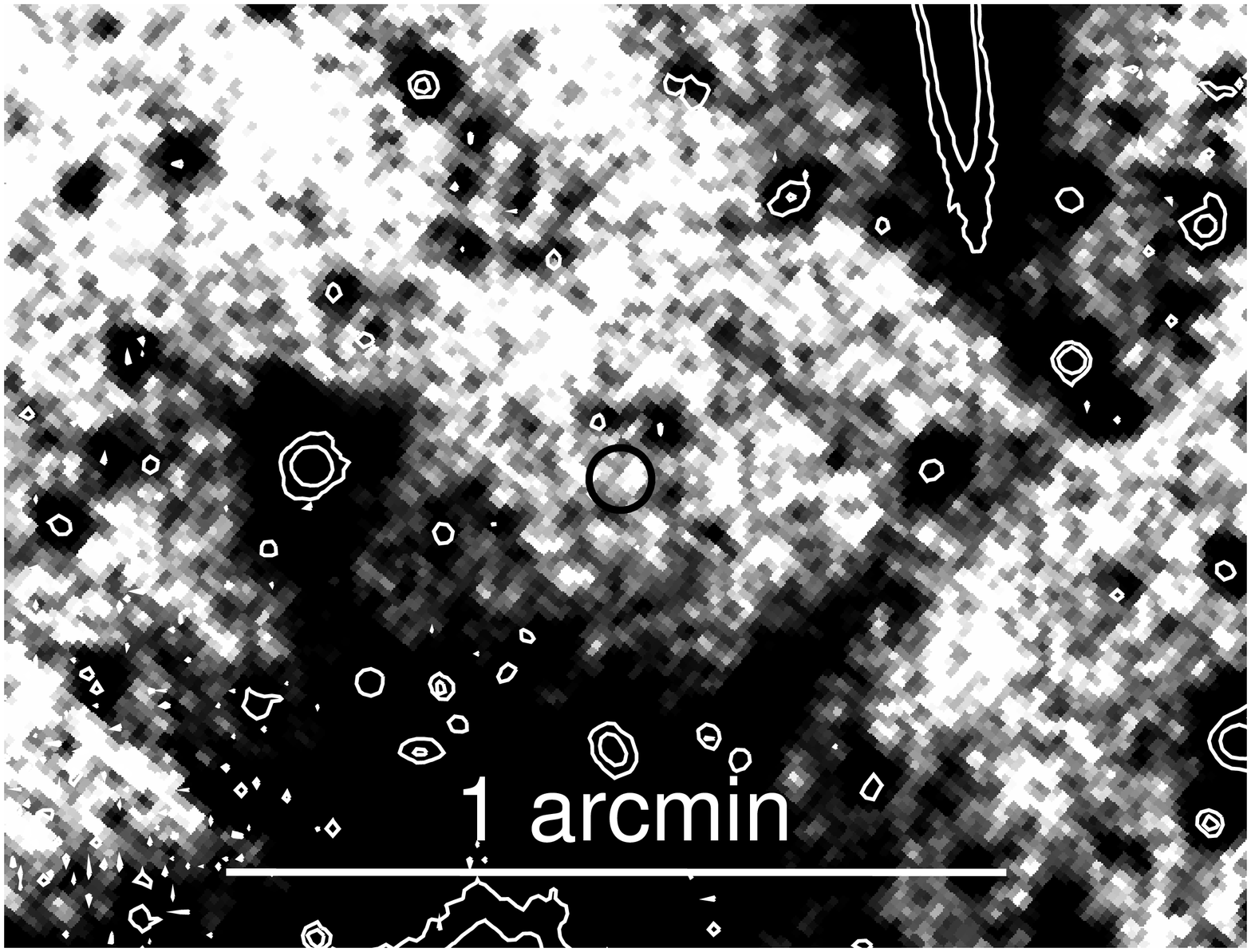}} \\
\caption{{\em Spitzer} IRAC2 ($4.5$\,$\mu$m) map around RX\,J0420.0--5022. The image is  $1\farcm{6} \times 1\farcm{2}$, North is up, East to the left. White contours are from our previous VLT $H$-band observations \citep{Posselt2009}. The INS position is marked with a black circle with a radius of 2 native IRAC pixels (4 image pixels, corresponding to $2\farcs{4}$). Note that the astrometric uncertainty for the expected RX\,J0420.0--5022 position is less than $2\arcsec$, according to constraints by \citet{Mignani2009}.
\label{j0420S}}
\end{center}
\end{figure}

There is no IR emission at the position of RX\,J0420.0--5022 in the {\em Spitzer} $4.5\mu$m IRAC image, see Figure~\ref{j0420S}. The aperture corrected $5\sigma$ upper limit is $F^{4.5\mu \rm m}_{5\sigma}=2.8$\,$\mu$Jy.\\

\begin{figure}[h]
\begin{center}
{\includegraphics[width=85mm]{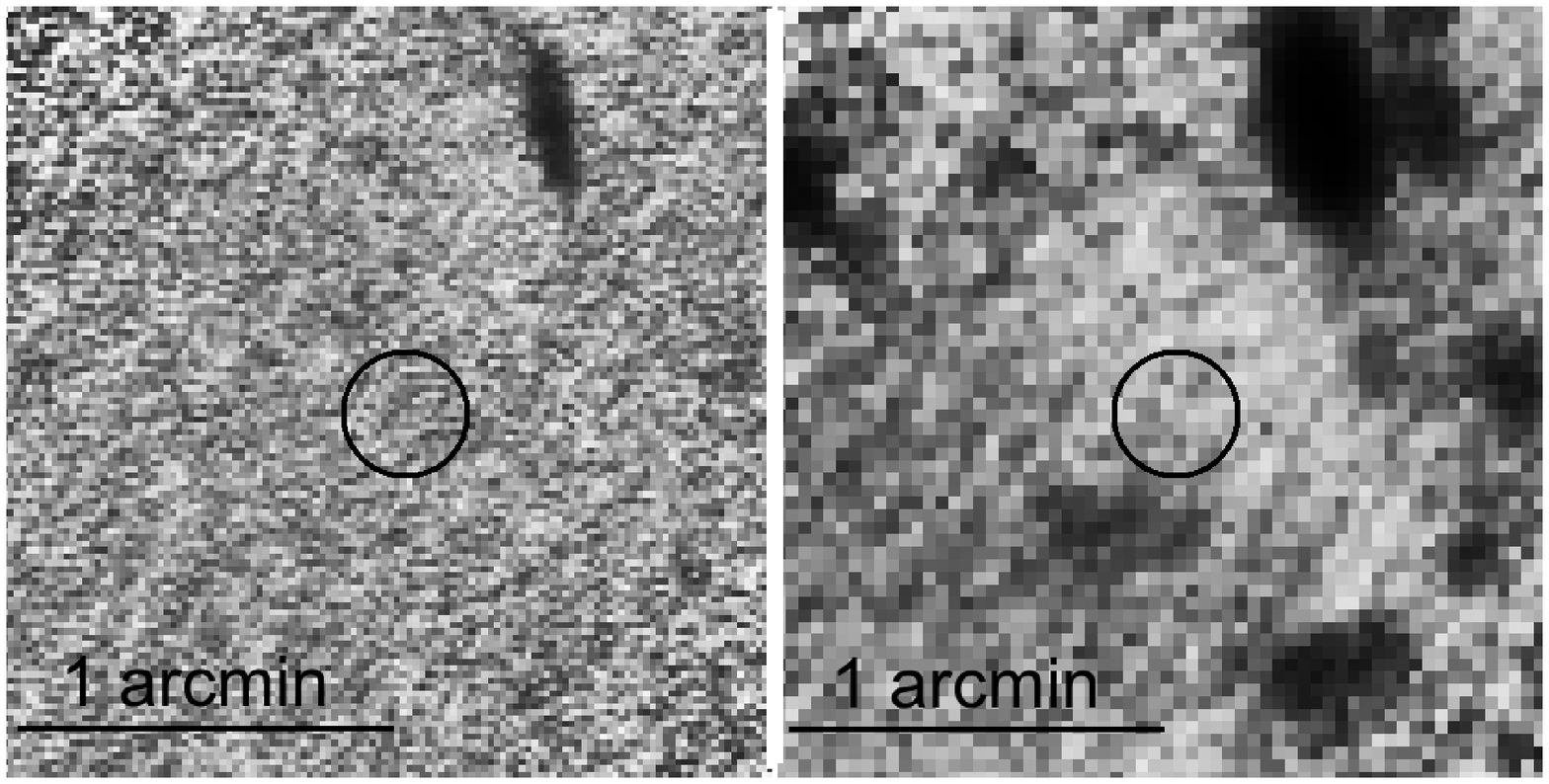}} \\
\caption{{\em Herschel} PACS maps around RX\,J0420.0--5022. The blue (60--85\,$\mu$m) and red (130--210\,$\mu$m) bands are shown in the left and right panels, respectively. Each map is $2\arcmin \times 2\arcmin$, North is up, East is to the left. The circle with a radius of $10\arcsec$ shows the NS position. 
\label{j0420H}}
\end{center}
\end{figure}

\begin{figure}[h]
\begin{center}
{\includegraphics[width=85mm]{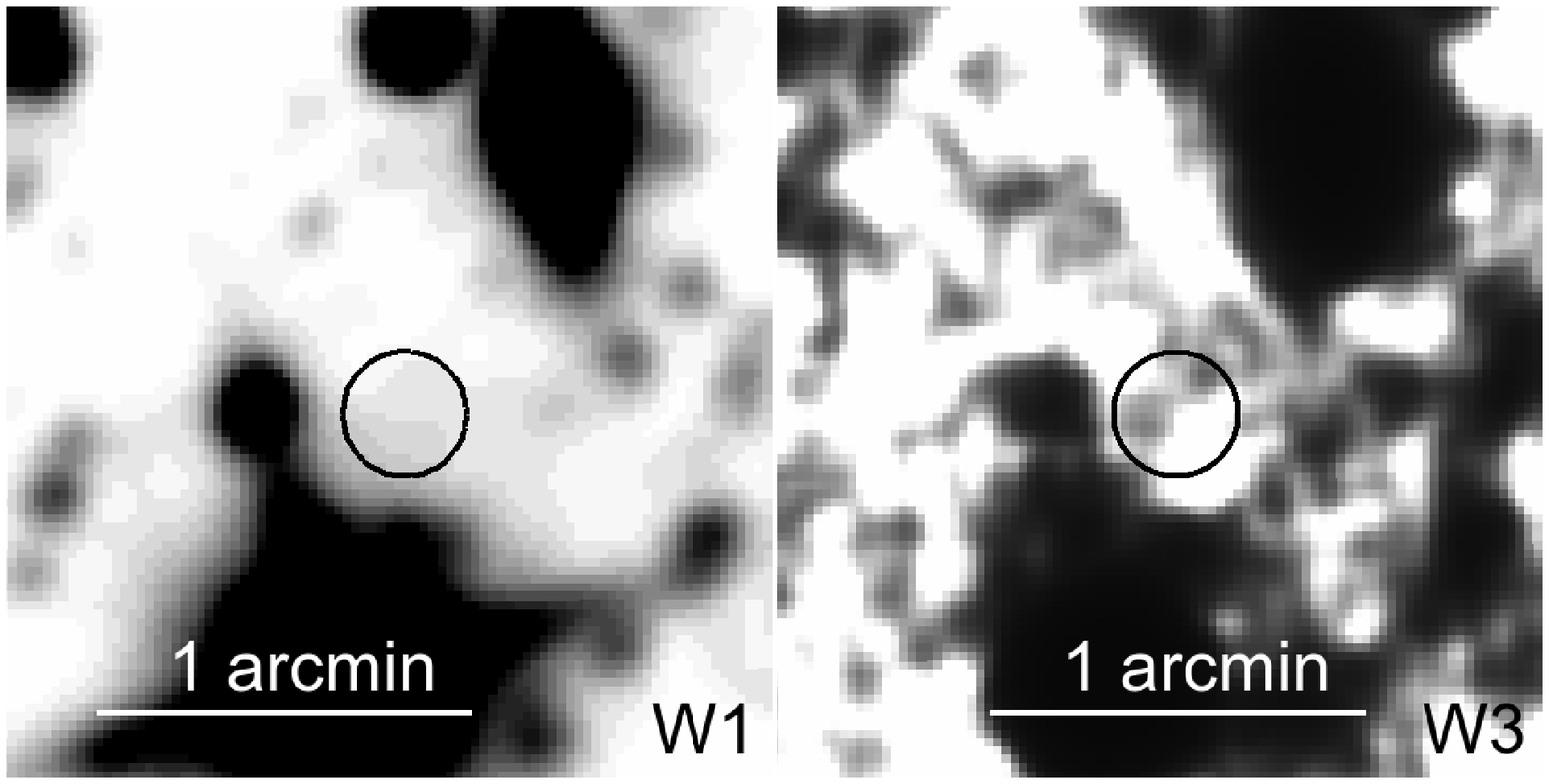}} \\
\caption{{\em WISE} W1 ($3.6$\,$\mu$m) and W3 (12\,$\mu$m) maps around the position of RX\,J0420.0--5022. The same field of view is shown as in the {\em Herschel} PACS maps of Figure~\ref{j0420H}.
\label{j0420W}}
\end{center}
\end{figure}

At the position of RX\,J0420.0--5022, we detect no source in the {\em Herschel} blue and red band (see Figure~\ref{j0420H}). Using apertures in different source-free regions and at the target position, we determined the $5\sigma$ upper flux limits at the INS position as $F^{\rm blue}_{5\sigma} = 4.5$\,mJy and $F^{\rm red}_{5\sigma} = 7$\,mJy. 
There are several common sources in the {\em Herschel} bands and the {\em WISE} bands W1 to W3. There is no noticeable astrometric shift between these bands. The  W1 and W3 images for the field around RX\,J0420.0--5022 are shown in Figure~\ref{j0420W}. There is very faint emission east and northwest of the INS in the W3 band, but no entry in the {\em WISE} source catalog at these positions. Slightly enhanced emission at the same positions might be suspected in the red band (Figure~\ref{j0420H}). However, these potential W3 sources are separated at least $5\farcs{8}$ from the expected INS position and are therefore unassociated to the INS.

\subsection{A.2. RX\,J0720.4--3125}
\label{sub0720}

\begin{figure}[h]
\begin{center}
{\includegraphics[width=85mm]{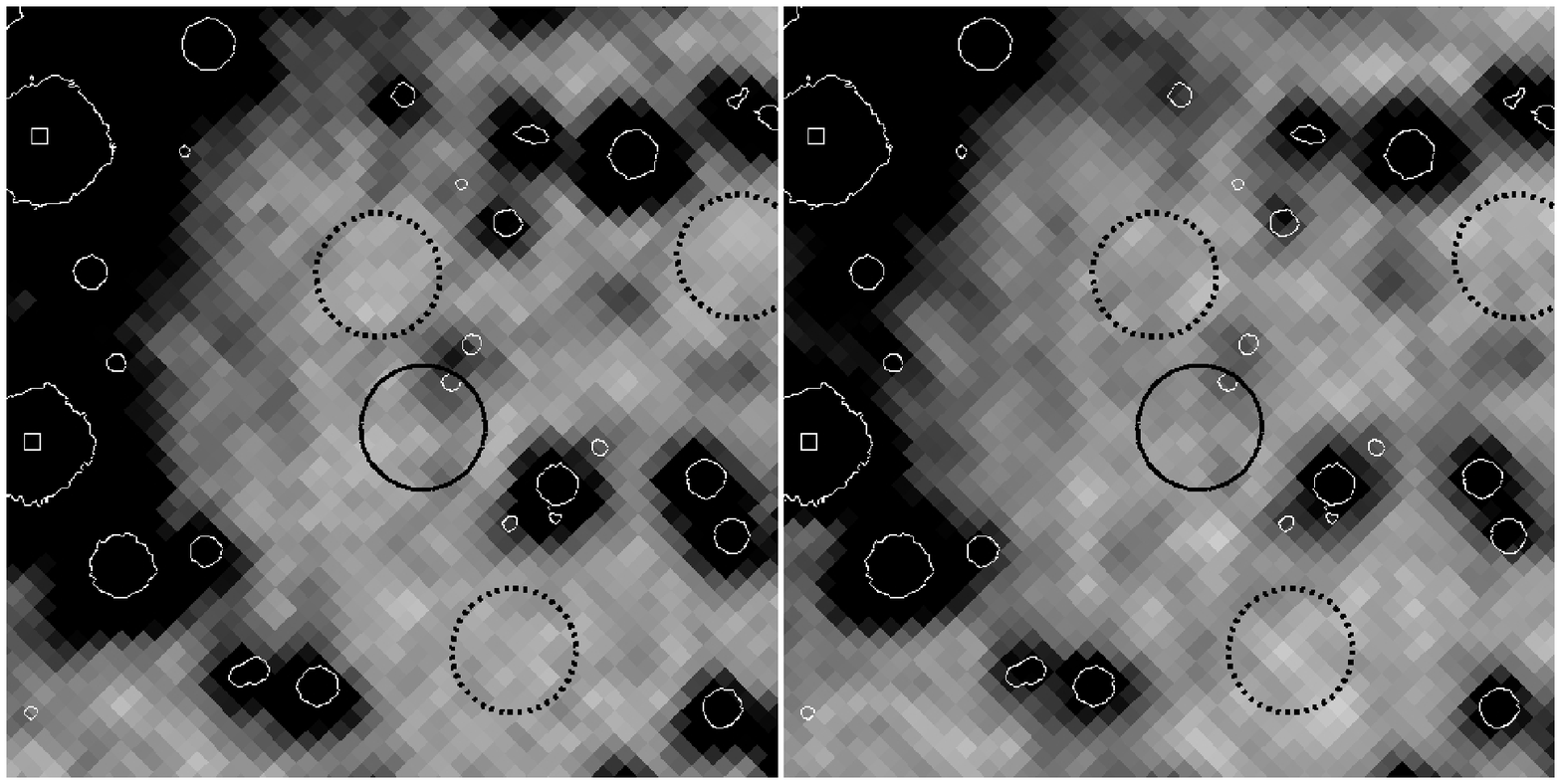}} \\
\caption{{\em Spitzer} $3.6$\,$\mu$m (left) and $4.5$\,$\mu$m (right) IRAC mosaic around RX\,J0720.4--3125.  Each map is $30\arcsec \times 30\arcsec$, North is up, East is to the left. The overlaid white contours are from the VLT ISAAC $H$-band image \citep{Posselt2009}, the small circles in two eastern sources mark 2MASS point source positions. The black circle with radius of $2\farcs{4}$ (2 native IRAC pixels) marks the position of RX\,J0720.4--3125 at the time of the {\em Spitzer} observations. The dashed circles indicate three examples of the ten apertures used to determine the background and noise in source-free image regions. 
\label{j0720S}}
\end{center}
\end{figure}

We first consider the {\sl{Spitzer}} observations. The IRAC mosaics are very crowded, and there appears to be a faint source close to the position of RX\,J0720.4--3125 (Figure~\ref{j0720S}).
The expected position of the INS at the time of the Spitzer observation, MJD 55582, is $\alpha = 7^h 20^m 24\fs{904} ;  \delta =  -31^\circ 25^\prime 49\farcs{63}$.
It is calculated using the latest, MJD=54477, optical position of the NS and its proper motion, $\mu_{\alpha} = -93.2 \pm 5.4$\,mas\,yr$^{-1}$ and $\mu_{\delta}=   48.6 \pm 5.1$\,mas\,yr$^{-1}$, listed by \citet{Eisenbeiss2010}. The radial positional uncertainty of the expected INS position at the time of the Spitzer observation is 62\,mas (90\% confidence level). The $1\sigma$ astrometric accuracy of the $3.6$\,$\mu$m and $4.5$\,$\mu$m mosaics is $0\farcs{27}$ and $0\farcs{28}$, respectively (Section~\ref{spitzerdatared}).
The faint IRAC $3.6$\,$\mu$m source closest to RX\,J0720.4--3125 seems to be a blended source and has a spatial separation of at least $1\farcs{3}$. From our previous VLT ISAAC $H$-band observations \citep{Posselt2009} two individual NIR sources are known close to the position of the faint IRAC source(s). From the spatial separation and the NIR identification of the faint IRAC source(s) we conclude that the IRAC source is not the counterpart of the INS. Thus, RX\,J0720.4--3125 is undetected at $3.6$\,$\mu$m and $4.5$\,$\mu$m.\\

To obtain an upper limit, we determine the background level and the standard deviation of 10 source-free apertures close to the source position. We use the residual mosaics (Section~\ref{spitzerdatared}) for these measurements. 
We chose \texttt{apex-qa} task parameters which result in a good removal of point sources in the immediate surrounding of the INS. The northern part of the faint blended source close to the INS is detected as individual point source and removed in both IRAC channels. A small flux enhancement remains in both channels at the position of the southern ISAAC source and slightly increases the derived upper limits, which are computed by the measured source aperture flux above the background plus five times the standard deviation of the 10 source-free apertures (Section~\ref{spitzerdatared}).  
The aperture-corrected $3.6$\,$\mu$m and the $4.5$\,$\mu$m upper flux limits at the position of RX\,J0720.4--3125 are $F^{3.6\mu \rm m}_{5\sigma}=4.9$\,$\mu$Jy and  $F^{4.5\mu \rm m}_{5\sigma}=4.5$\,$\mu$Jy, respectively.\\

\begin{figure}[h]
\begin{center}
{\includegraphics[width=85mm]{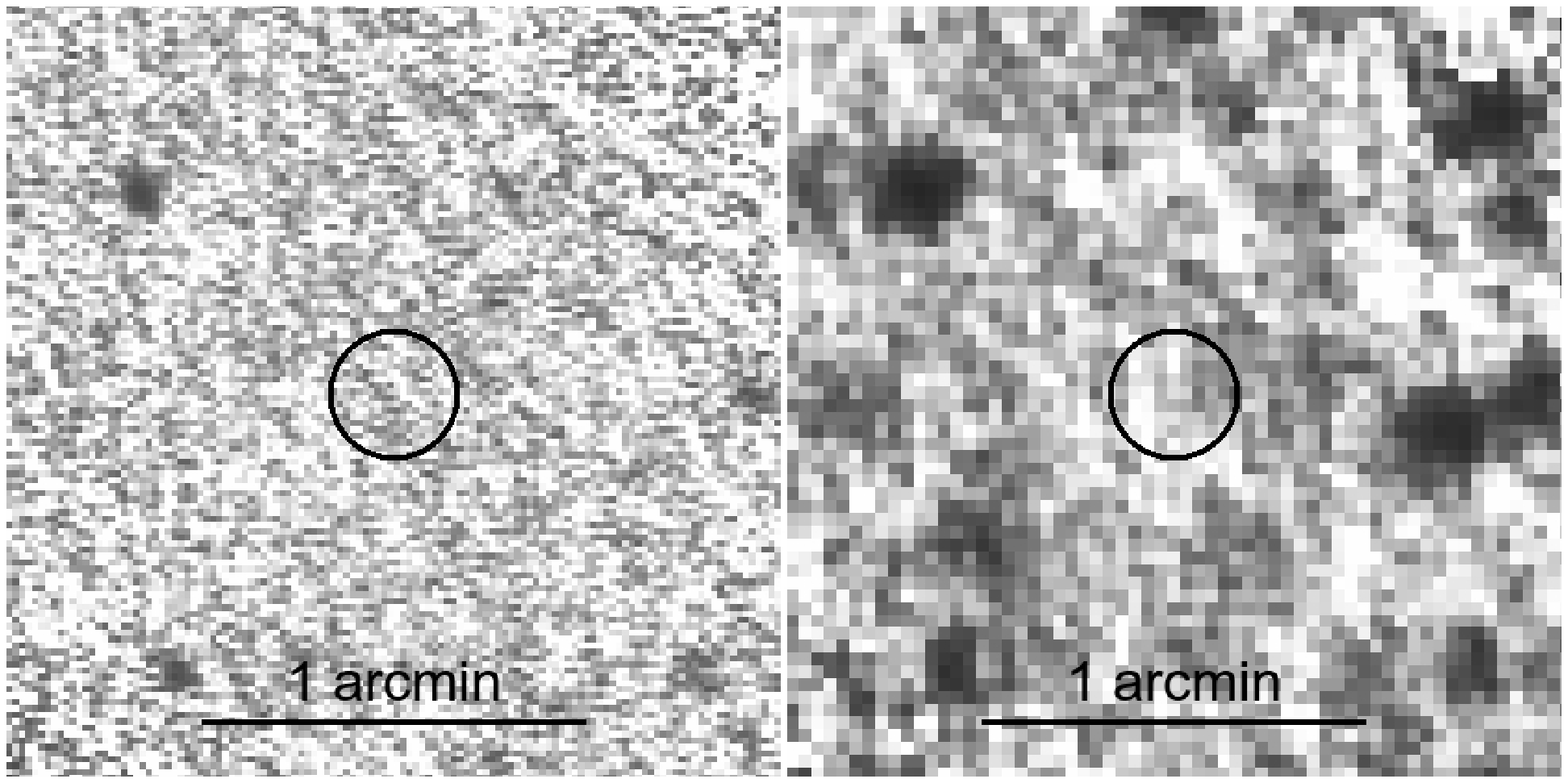}} \\
\caption{{\em Herschel} PACS maps around RX\,J0720.4--3125. The blue (60--85\,$\mu$m) and red (130--210\,$\mu$m) bands are shown in the left and right panels, respectively. Each map is $2\arcmin \times 2\arcmin$, North is up, East is to the left. The circle with a radius of $10\arcsec$ shows the XTINS position. 
\label{j0720H}}
\end{center}
\end{figure}

There is no obvious infrared source in the {\em Herschel} blue and red band at the position of RX\,J0720.4--3125 (see Figure~\ref{j0720H}).
The closest possible source in the red band is at least $6\arcsec$ from the INS position.
Using apertures in different source-free regions and at the target position, we determine the $5\sigma$ upper flux limits for the INS to be $F^{\rm blue}_{5\sigma} = 5.2$\,mJy and $F^{\rm red}_{5\sigma} = 9.4$\,mJy. \\

\begin{figure}[h!]
\begin{center}
{\includegraphics[height=50mm, bb=297 20 575 301, clip]{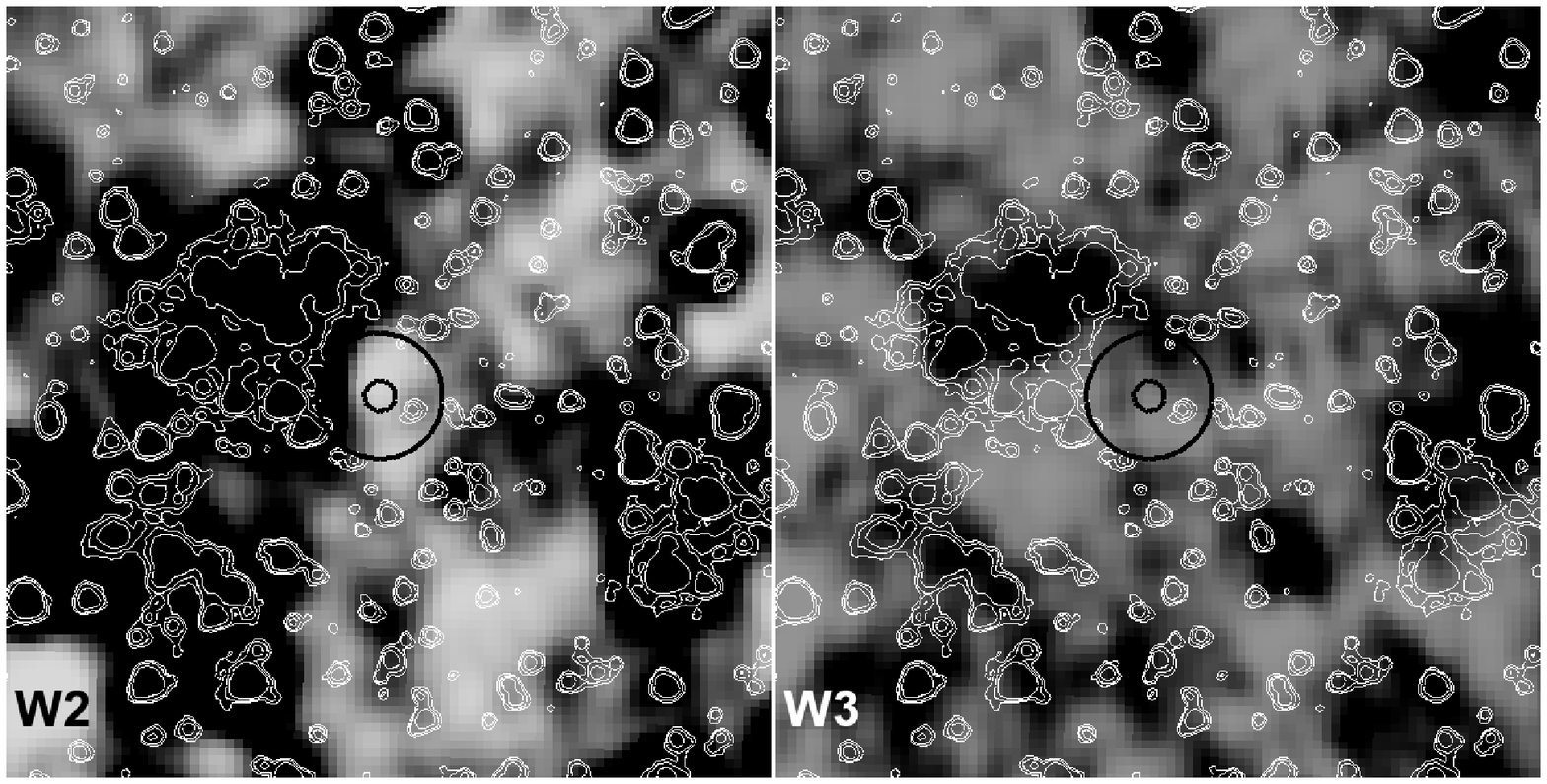}} \\
\caption{W3 (12\,$\mu$m) map around the position of RX\,J0720.4--3125. The  field of view is $2\arcmin \times 2\arcmin$, north is up, east is to the left. The white overlay contours are the {\sl{Spitzer}}  $4.5$\,$\mu$m IRAC contours. The black circles mark the position of RX\,J0720.4--3125, the small one is the same as indicated in the {\sl{Spitzer}} image (Fig.~\ref{j0720S}, radius $2\farcs{4}$), the larger one has a radius of $10\arcsec$ as in the {\em Herschel} PACS maps (Fig.~\ref{j0720H}).
\label{j0720W}}
\end{center}
\end{figure}

The spatial resolution and sensitivity of the two {\sl{Spitzer}} IRAC images are superior to the first two channels of the {\em WISE} data, but the W3 and W4 band provide new wavelength coverage. Nothing is seen in W3 or W4 at the position of the INS. Because of the uncertainty in the W4 astrometry, only W3 is shown in Figure~\ref{j0720W}. The closest, very faint 12\,$\mu$m source in the east of the INS has a spatial separation from the INS of $\sim 5\arcsec$. Given the good astrometric compliance of major W3 sources with {\sl{Spitzer}} IRAC sources, this faint source is unlikely to be the INS counterpart. 

\subsection{A.3. RX\,J1308.6+2127} 
\label{sub1308}
There is no IR emission at the position of RX\,J1308.6+2127 in the {\em Spitzer} $4.5\mu$m IRAC image, see Figure~\ref{j1308s}. The aperture corrected $5\sigma$ upper limit is $F^{4.5\mu \rm m}_{5\sigma}=1.6$\,$\mu$Jy.\\

\begin{figure}[h]
\begin{center}
{\includegraphics[width=85mm]{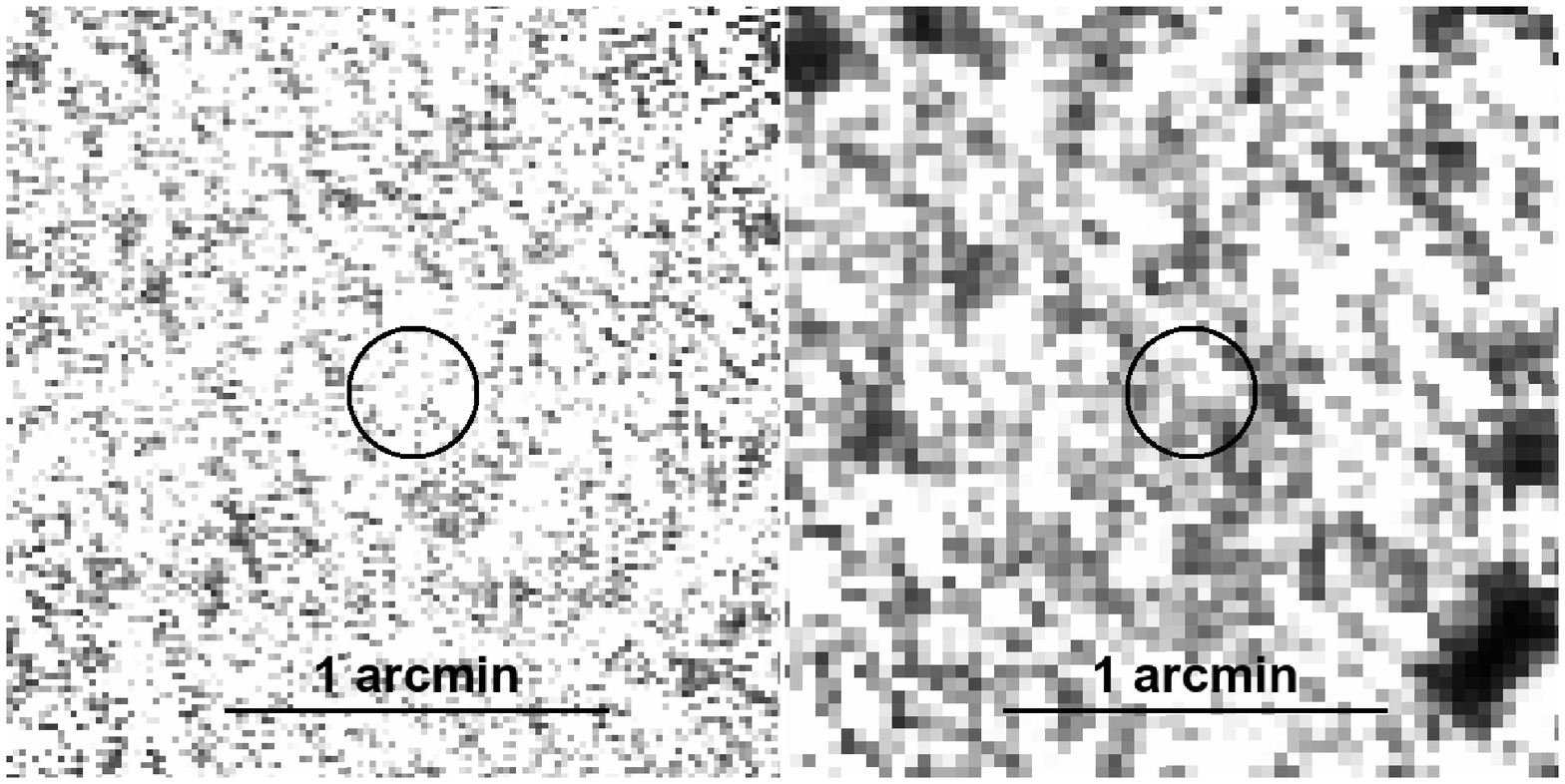}} \\
\caption{{\em Herschel} PACS maps around RX\,J1308.6+2127. The blue (60--85\,$\mu$m) and red (130--210\,$\mu$m) bands are shown in the left and right panels, respectively. Each map is $2\arcmin \times 2\arcmin$, North is up, East is to the left. The circle with a radius of $10\arcsec$ shows the NS  position. \label{j1308H}}
\end{center}
\end{figure}

\begin{figure}[h]
\begin{center}
{\includegraphics[width=85mm]{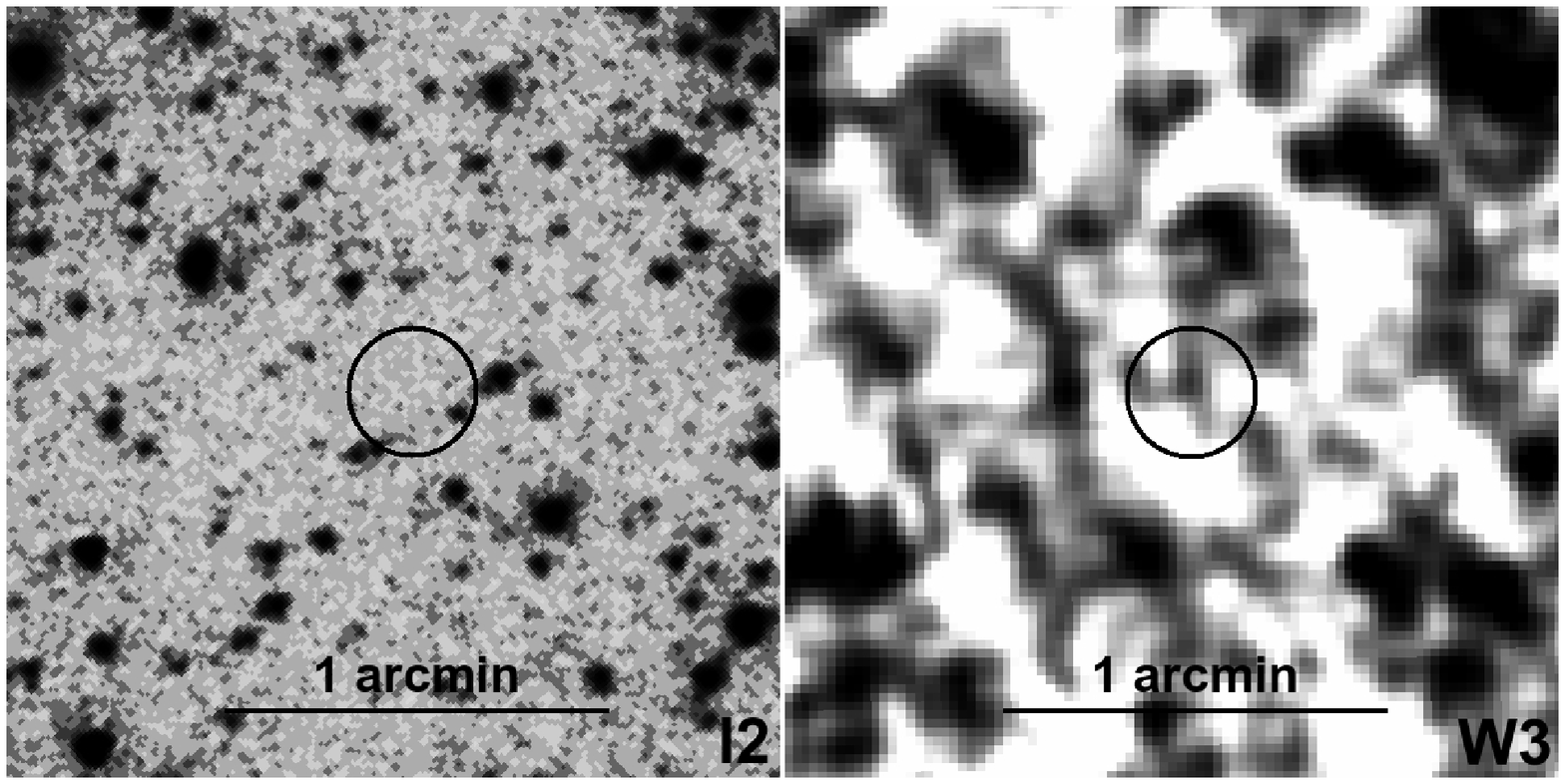}} \\
\caption{{\em Spitzer} IRAC2 ($4.5$\,$\mu$m, left) and {\em WISE} W3 (12\,$\mu$m, right) maps around the position of RX\,J1308.6+2127. The same field of view is shown as in the {\em Herschel} PACS maps of Figure~\ref{j1308H}.
\label{j1308s}}
\end{center}
\end{figure}

In the {\em Herschel} PACS blue band, there is no significant emission around the position of RX\,J1308.6+2127, see Figure~\ref{j1308H}. Using apertures in different source-free regions and at the target position we determine the $5\sigma$ upper flux limit for the INS as $F^{\rm blue}_{5\sigma} = 1.7$\,mJy. 
In the {\em Herschel} PACS red band, there is no significant emission at the position of RX\,J1308.6+2127, but there is faint emission southwest and west of it (Figure~\ref{j1308H}). The spatial separation of this faint emission and the XTINS position is $10\arcsec$. 
The positional uncertainty of the NS at the 90\% confidence level in the {\em Herschel} observation is smaller than $0\farcs{7}$ with account of the uncertainty of the {\em Chandra} position \citep{Kaplan2002,Hambaryan2002} and the uncertainty of the proper motion \citep{Motch2009}. 
We checked for common sources in the {\em Herschel} observation, 2MASS point source catalog and {\em Spitzer} IRAC $4.5$\,$\mu$m data. Based on this comparison, we confirm that the absolute {\em Herschel} pointing accuracy is better than $1\farcs{5}$ for this observation as expected from the works by the {\em Herschel} calibration and HIPE teams (see above). Thus, all the faint sources in the {\em Herschel} PACS red band can be excluded as INS counterparts.
Using apertures in different source-free regions and at the target position, we determine the $5\sigma$ upper flux limit in the red band for the INS as  $F^{\rm red}_{5\sigma} = 5.2$\,mJy.\\ 

We compared the {\em WISE} W1-W3 data to the {\em Spitzer} IRAC $4.5$\,$\mu$m data. There is no noticeable astrometric shift between {\em Spitzer} IRAC $4.5$\,$\mu$m and {\em WISE} W1 and W2. For {\em WISE} W3 there are only few clear common sources, three of them are 2MASS point sources. There is no noticeable astrometric shift for these sources. Interestingly, there is faint emission  in {\em WISE} W3 at a position consistent with the INS position within positional accuracy, see Figure~\ref{j1308s}. There is no source reported in the {\em WISE} All-Sky Source Catalog at this position, and no such emission is seen in any other {\em WISE} band, the {\em Spitzer} IRAC $4.5$\,$\mu$m or the {\em Herschel} PACS bands. 
Due to its faintness the W3 emission can be a noise feature. Confirmation by similar observations are necessary before any conclusions regarding the INS can be drawn.
In principle, emission at such wavelengths is particularly interesting, because -- if real -- it could indicate potential silicate emission in the $8-12$\,$\mu$m region reminiscent of white dwarf debris disks (e.g., in the case of the spectacularly debris-polluted white dwarf GD 362, one-third of the debris disk emission is carried by a strong $10\mu$m silicate emission feature; \citealt{Jura2007}).

\subsection{A.4. RX\,J1605.3+3249} 
\label{sub1605}
\begin{figure}[h]
\begin{center}
{\includegraphics[width=55mm]{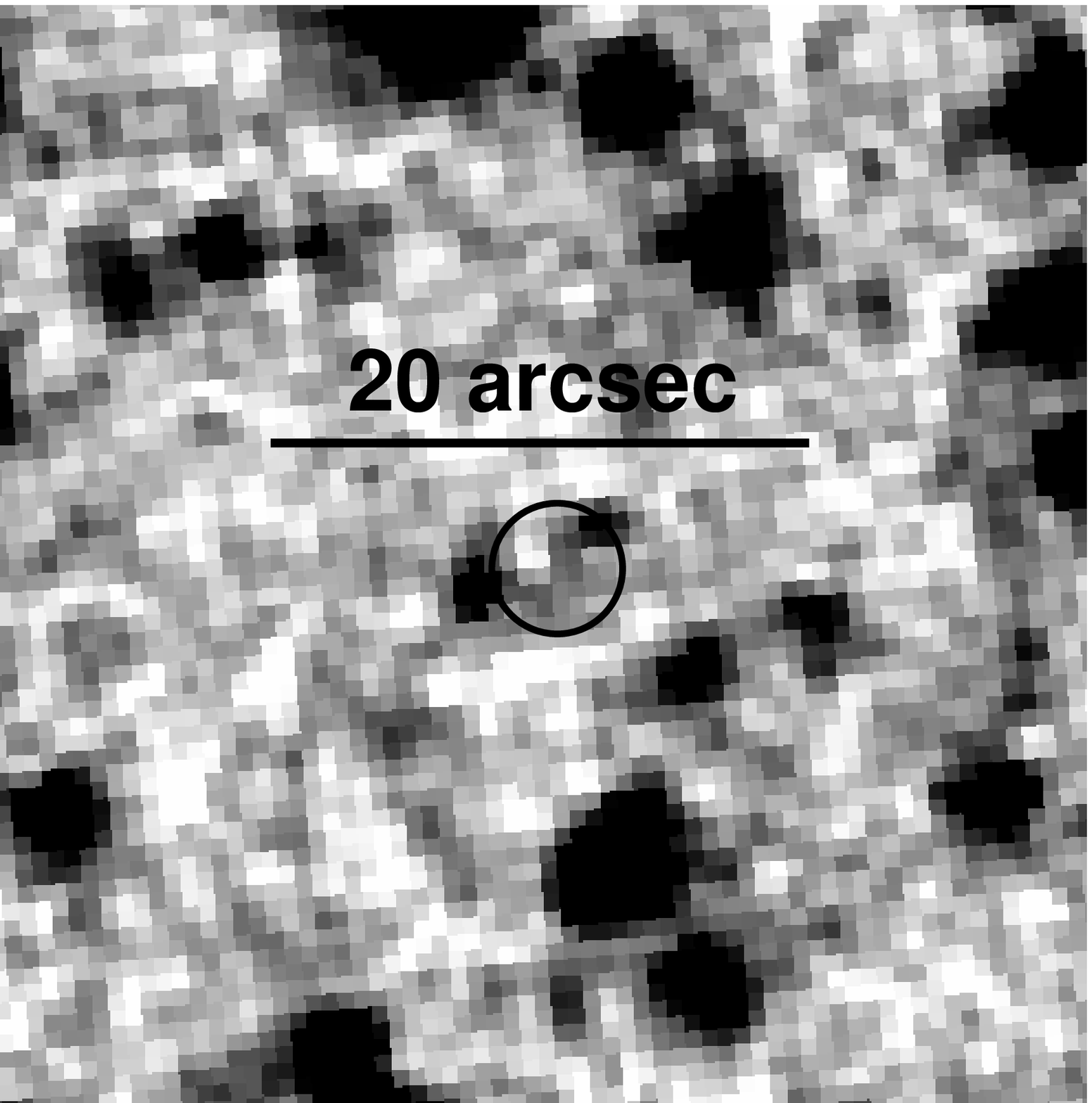}} \\
\caption{{\em Spitzer} IRAC2 ($4.5$\,$\mu$m) map around RX\,J1605.3+3249 ( $40\arcsec \times 40\arcsec$, North is up, East is to the left). The circle with a radius of $2.4\arcsec$ shows the expected INS position at the time of the {\em Spitzer} observations. 
\label{j1605S}}
\end{center}
\end{figure}

There is very faint IR emission at the position of RX\,J1605.3+3249 in the {\em Spitzer} $4.5\mu$m IRAC image, see Figure~\ref{j1605S}. However, the INS position is located between two IR sources whose PRFs affect any measured flux at the INS position. We found that the faint $4.5\mu$m emission has a significance of only $1.3\sigma$ if we employ PRF subtraction of the two sources in the vicinity ($2.9\sigma$ otherwise). Hence, there is no significant emission at $4.5\mu$m from the INS.
The aperture corrected $5\sigma$ upper limit is $F^{4.5\mu \rm m}_{5\sigma}=3.6$\,$\mu$Jy.\\

\begin{figure}[h]
\begin{center}
{\includegraphics[width=85mm]{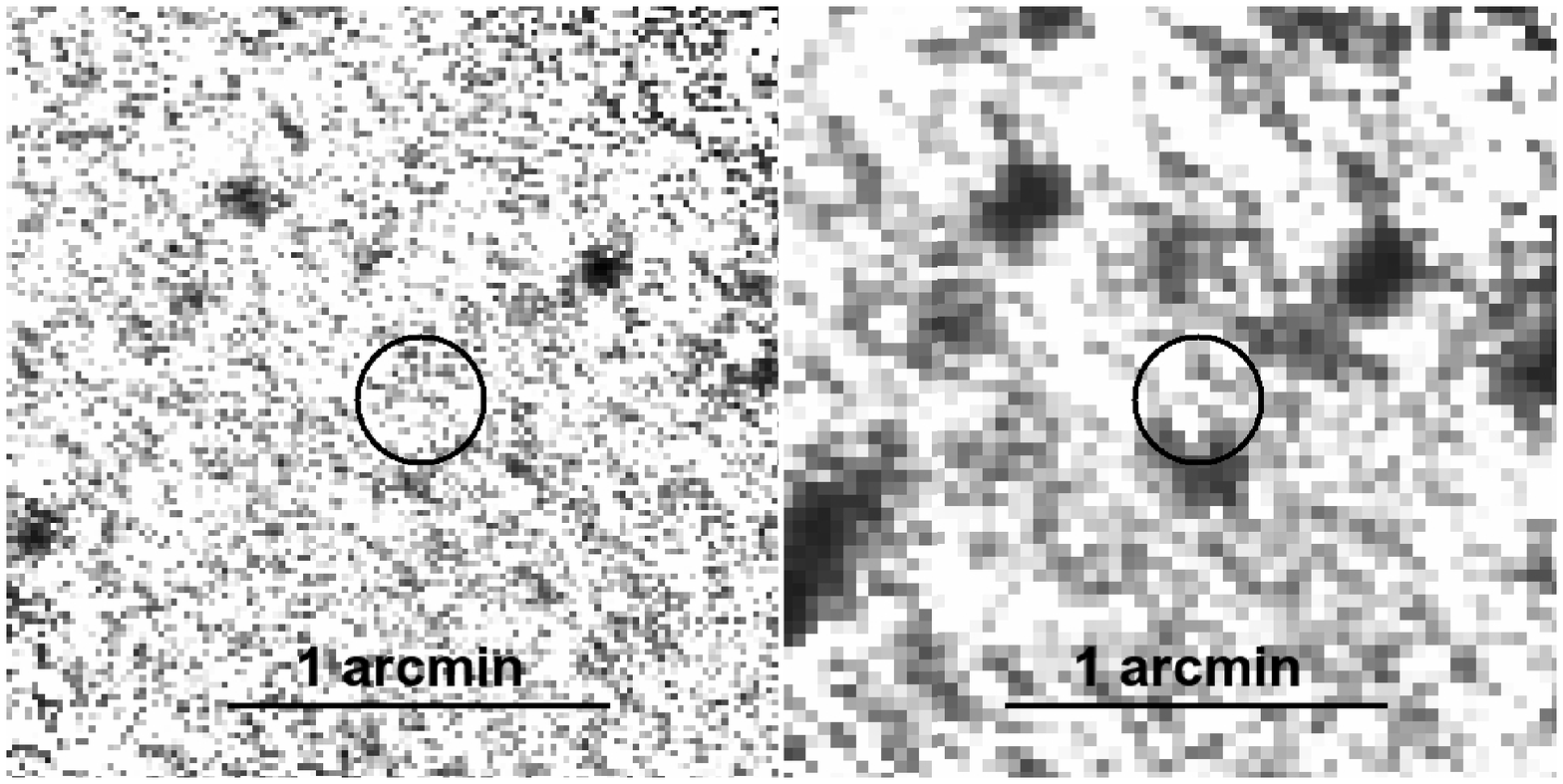}} \\
\caption{{\em Herschel} PACS maps around RX\,J1605.3+3249. The blue (60--85\,$\mu$m) and red (130--210\,$\mu$m) bands are shown in the left and right panels, respectively. Each map is $2\arcmin \times 2\arcmin$, North is up, East is to the left. The circle with a radius of $10\arcsec$ shows the NS  position. 
\label{j1605H}}
\end{center}
\end{figure}

\begin{figure}[h]
\begin{center}
{\includegraphics[width=85mm]{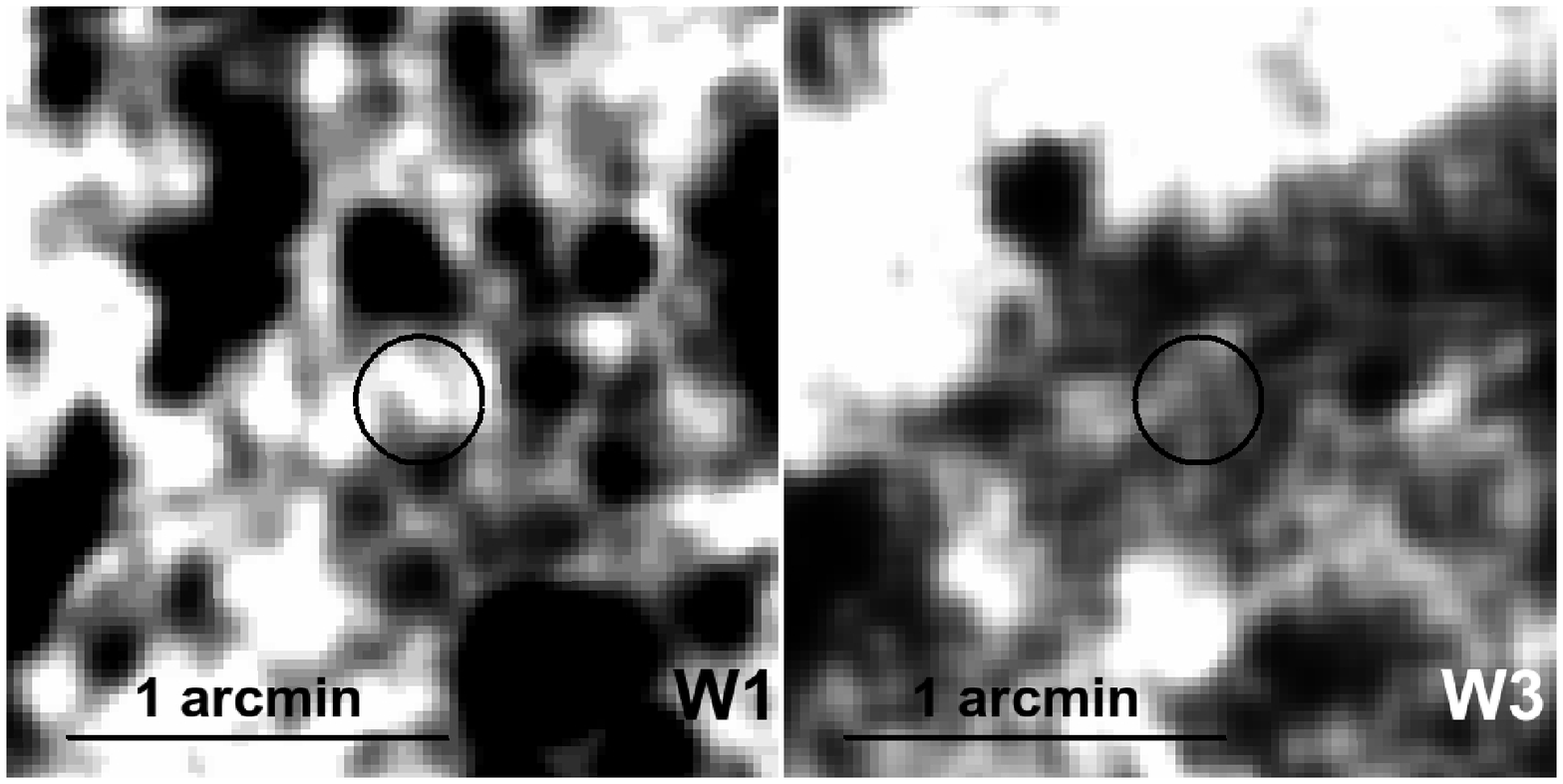}} \\
\caption{{\em WISE} W1 ($3.6$\,$\mu$m) and W3 (12\,$\mu$m) maps around the position of RX\,J1605.3+3249. The same field of view is shown as in the {\em Herschel} PACS maps of Figure~\ref{j1605H}.
\label{j1605W}}
\end{center}
\end{figure}
There are several common sources in the {\em Herschel} bands and the {\em WISE} bands W1 to W3. There is no noticeable astrometric shift between these bands. There is no source in the {\em Herschel} blue and red bands at the position of RX\,J1605.3+3249 (see Figure~\ref{j1605H}). North of the INS there is a faint source in the red band which is at least $5\farcs{6}$ separated from the expected INS position and is therefore unassociated to the INS.
Using apertures in different source-free regions and at the target position, we determine the $5\sigma$ upper limits as $F^{\rm blue}_{5\sigma} = 6.1$\,mJy and $F^{\rm red}_{5\sigma} = 12.2$\,mJy. 
The {\em WISE}  W1 and W3 images for the field around RX\,J1605.3+3249 are shown in Figure~\ref{j1605W}. While there is large scale extended emission in W3, no point source is seen at position of RX\,J1605.3+3249 in any of the {\em WISE} bands.

\subsection{A.5. RX\,J1856.5--3754}
\label{sub1856}
\begin{figure}[h]
\begin{center}
{\includegraphics[width=85mm]{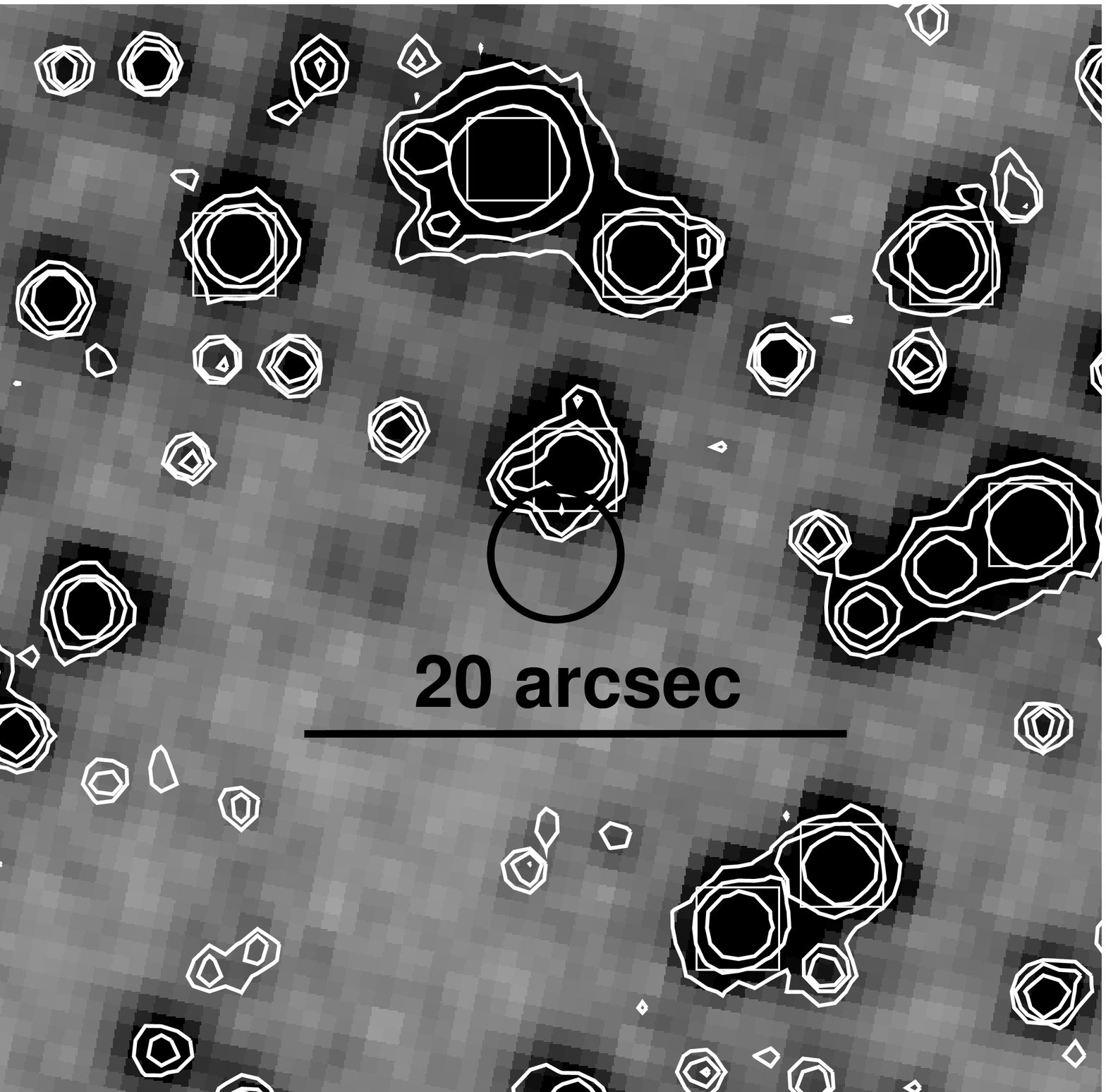}} \\
\caption{{\em Spitzer} IRAC2 ($4.5$\,$\mu$m) map around RX\,J1856.5--3754. The image is $40\arcsec \times 40\arcsec$, North is up, East is to the left. The circle with a radius of $2.4\arcsec$ shows the expected INS position at the time of the {\em Spitzer} observations. The overplotted white contours are from the VLT ISAAC $H$-band image \citep{Posselt2009}, the white boxes indicate 2MASS point sources.
\label{j1856S}}
\end{center}
\end{figure}
The positions of RX\,J1856.5--3754 at the time of the {\em Spitzer} and {\em Herschel} observations were estimated from the position and proper motion values listed by \citet{Walter2010}. The accuracy of the determined positions are better than $0\farcs{1}$. 
The astrometry of the {\em Spitzer} IRAC2 ($4.5$\,$\mu$m) image of RX\,J1856.5--3754 is well aligned with the astrometry of previous $H$-band observations (Figure~\ref{j1856S}). There is no prominent emission centered at the expected INS position, but there is strong IR emission north of RX\,J1856.5--3754, starting at $\approx 1\arcsec$ from the INS. From the NIR we know this source consists of at least four individual sources which are, however, not discernable with  {\em Spitzer}. We were not able to completely remove the IR emission of these northern sources by PRF fitting. Thus, our subsequently derived aperture-corrected $5\sigma$ upper limit for RX\,J1856.5--3754, $F^{4.5\mu \rm m}_{5\sigma}=11.6$\,$\mu$Jy, is a very conservative estimate.\\

\begin{figure}[h]
\begin{center}
{\includegraphics[width=85mm]{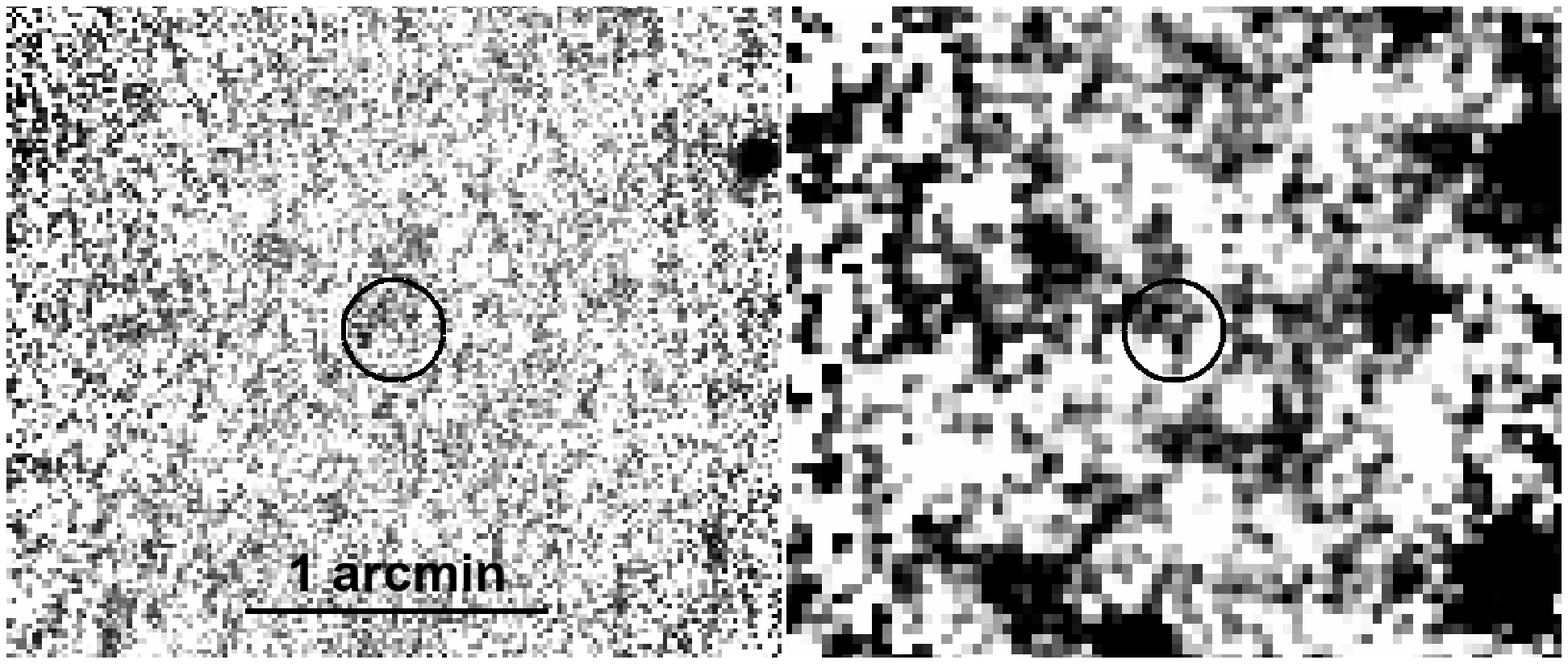}} \\
\caption{{\em Herschel} PACS maps around RX\,J1856.5--3754. The blue (60--85\,$\mu$m) and red (130--210\,$\mu$m) bands are shown in the left and right panels, respectively. North is up, East is to the left. The circle with a radius of $10\arcsec$ shows the NS  position. 
\label{j1856H}}
\end{center}
\end{figure}

Comparing {\em Herschel} PACS and {\em WISE} sources we were not able to identify enough unique common sources to check the astrometry of the  {\em Herschel} observations. Therefore, we assume the astrometric error to be the absolute pointing error of $\approx 2\farcs{4}$. This value was listed by the {\em Herschel} calibration team as average value for a time window which includes our observations\footnote{herschel.esac.esa.int/twiki/bin/view/Public/SummaryPointing}.
In the {\em Herschel} blue band, there is very faint emission at $\approx 4\farcs{6}$ north-east from the INS position.
In the {\em Herschel} red band, the INS position is located within a larger faint emission region which probably consists of several sources and extends to the north-east and north (see Figure~\ref{j1856H}).
We cannot exclude that the INS contributes to the faint PACS $160$\,$\mu$m band emission. However, it appears likely that most of that emission actually comes from the same NIR sources which are also detected by {\em Spitzer}.    
We measured the flux at the INS position in an aperture with radius $5\arcsec$. The aperture-corrected flux is $F^{\rm red}=3.1 $\,mJy, which corresponds to $\approx 3\sigma$ above the background. The nominal $5\sigma$ flux limit for the position of RX\,J1856.5--3754 in the red band is $F^{\rm red}_{5\sigma}=7.8 $\,mJy.
In the blue band, we used apertures in different source-free regions and at the target position to determine the $5\sigma$ upper flux limit at the INS position as $F^{\rm blue}_{5\sigma} = 7.7$\,mJy.\\ 

\begin{figure}[h]
\begin{center}
{\includegraphics[width=95mm]{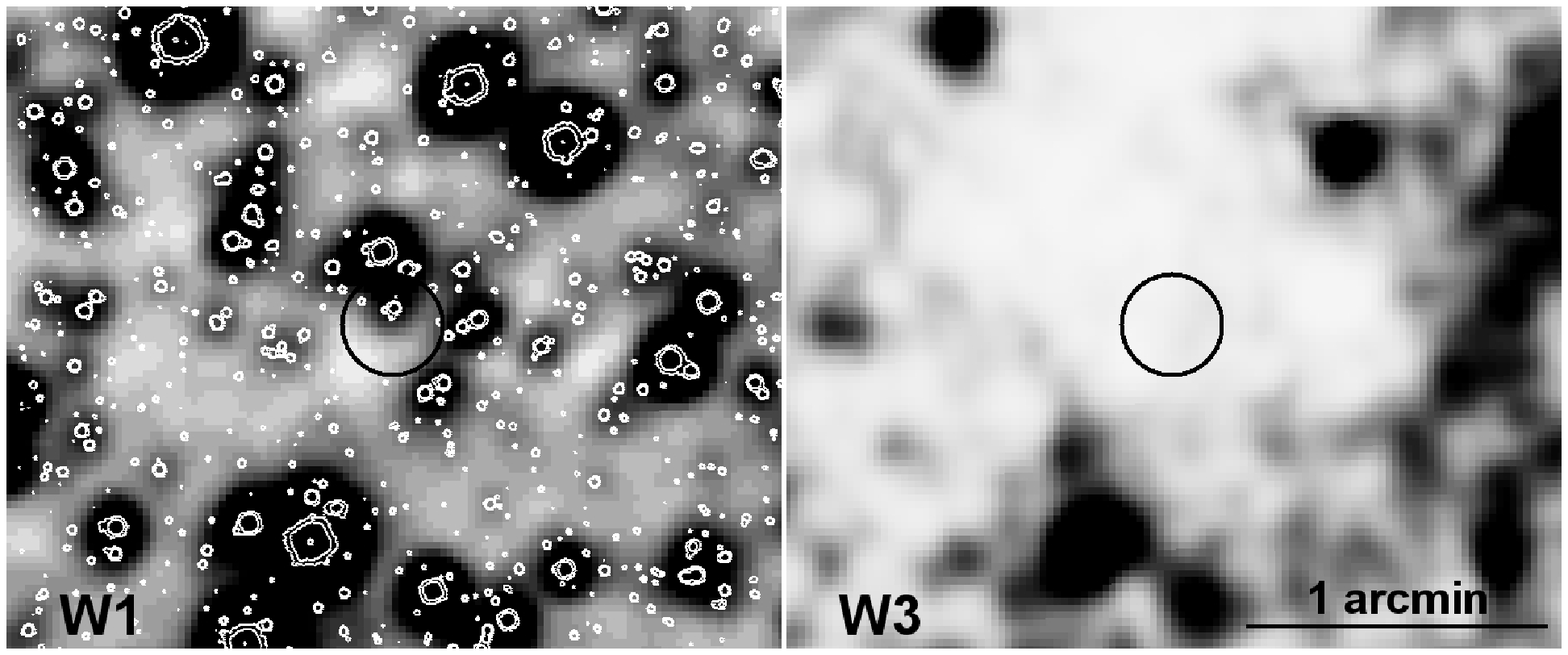}} \\
\caption{{\em WISE} W1 ($3.6$\,$\mu$m) and W3 (12\,$\mu$m) maps around the position of RX\,J1856.5--3754. The same field of view is shown as in the {\em Herschel} PACS maps of Figure~\ref{j1856H}.The white contours in the W1 image are from the VLT ISAAC $H$-band image \citep{Posselt2009}.
\label{j1856W}}
\end{center}
\end{figure}

Comparing sources in the {\em WISE} W1-W3 bands images with $H$-band sources from a deep VLT observation \citep{Posselt2009}, there is no apparent astrometric shift between these bands (see Figure~\ref{j1856W}). The IR sources observed with W1 and W2  correspond to the ones seen with {\em Spitzer} IRAC $4.5$\,$\mu$m. Nothing is detected around the INS position in W3.

\subsection{A.6. PSR\,J1848--1952} 
\label{psr1848}
The position of PSR\,J1848--1952 at MJD $48695$ is known with a $2\sigma$ accuracy of $0\farcs{6}$ in right ascension, but $7\arcsec$ in declination \citep{Hobbs2004}. The proper motion is unknown for this pulsar. 
PSR\,J1848--1952 was recently observed for 38\,ksec with  XMM-$Newton$ (obsid:0653300101, PI Kaspi). Unfortunately, the pulsar was not detected \citep{Olausen2012}, the closest X-ray source to the pulsar radio position has an angular separation of $\approx 1\farcm{65}$. 
Thus, the X-ray observations cannot be used to determine the current position of PSR\,J1848--1952 with higher accuracy.
\citet{Hobbs2004} gave the following proper motion values $\mu_{\alpha} \cos \delta =    39(157)$\,mas yr$^{-1}$ and $\mu_{\delta}=   -1200(1900)$ \,mas yr$^{-1}$. While the error in declination is too large to be used for a reasonable upper limit on the expected pulsar proper motion, we use the error in right ascension to determine the maximal angular separation of the possible pulsar position in this direction at the time of the {\em Herschel} observation as $3\farcs{1}$ ($2\sigma$). 
The mean 2D speed of pulsars is $\approx 250$\,km\,s$^{-1}$, and PSR\,B2224+64 has the highest inferred 2D speed of $\approx 1600$\,km\,s$^{-1} $\citep{Hobbs2005}. We assume this most extreme case to determine the maximal expected shift in declination for PSR\,J1848--1952.
Its DM$=18.23$\,cm$^{-3}$\,pc  translates into a distance of 750\,pc using the NE2001 model for the Galactic distribution of free electrons \citep{ne2001}. We estimate the upper limit of the proper motion of PSR\,J1848--1952 as $0\farcs{45}$\,yr$^{-1}$. Thus, at the time of the {\em Herschel} observation, the maximal 
expected shift is $9\arcsec$. We assume this value as upper limit on the proper motion in the direction of declination.\\
\begin{figure}[h]
\begin{center}
{\includegraphics[width=85mm]{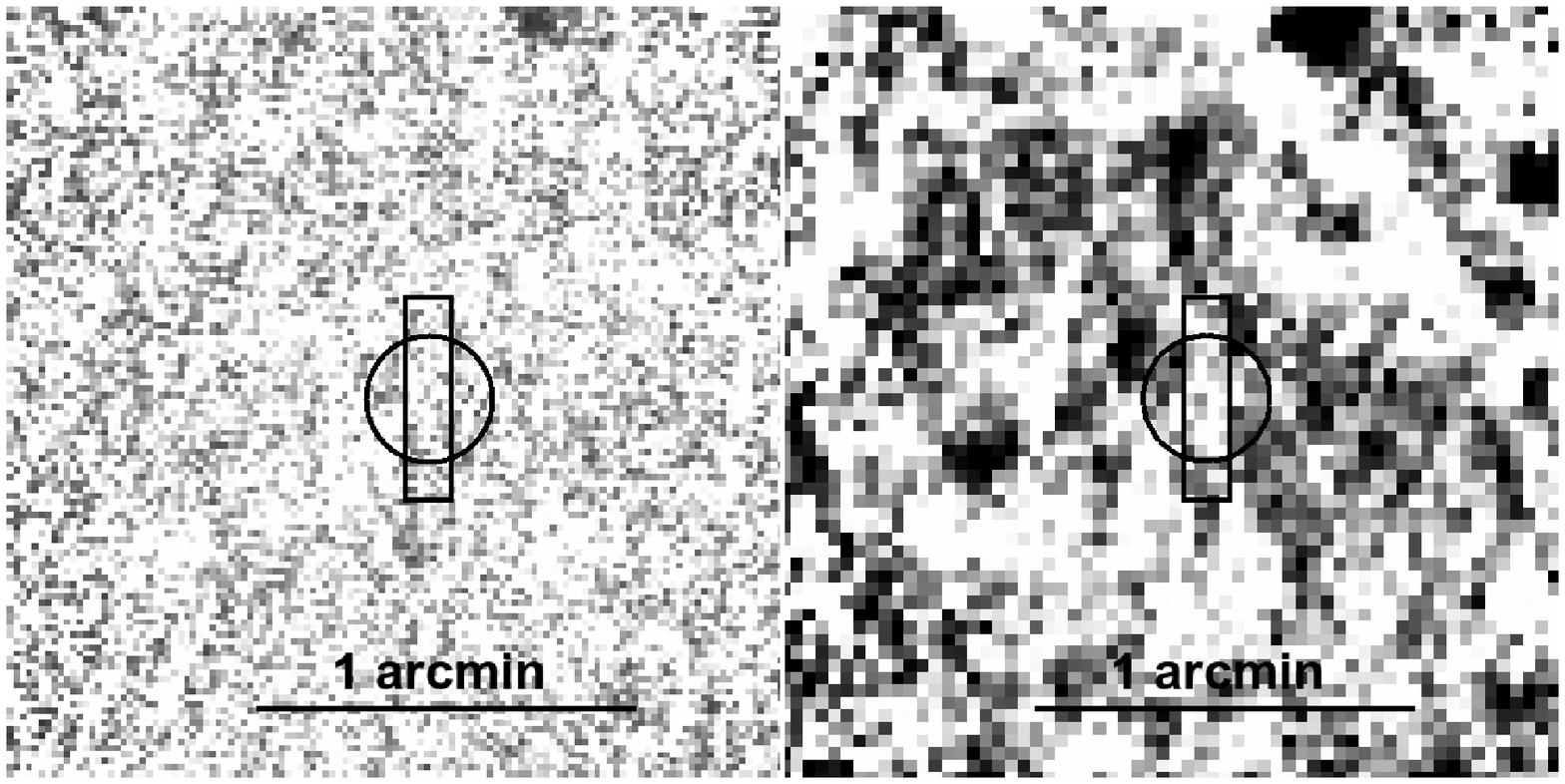}} \\
\caption{{\em Herschel} PACS maps around PSR\,J1848--1952. The blue (60--85\,$\mu$m) and red (130--210\,$\mu$m) bands are shown in the left and right panels, respectively. Each map is $2\arcmin \times 2\arcmin$, North is up, East is to the left. The circle with a radius of $10\arcsec$ marks the pulsar position, the $7\farcs{4} \times 32\arcsec$ box shows the upper limits of the unknown proper motion together with the radio position uncertainty (see text). 
\label{j1848H}}
\end{center}
\end{figure}

\begin{figure}[h]
\begin{center}
{\includegraphics[width=85mm]{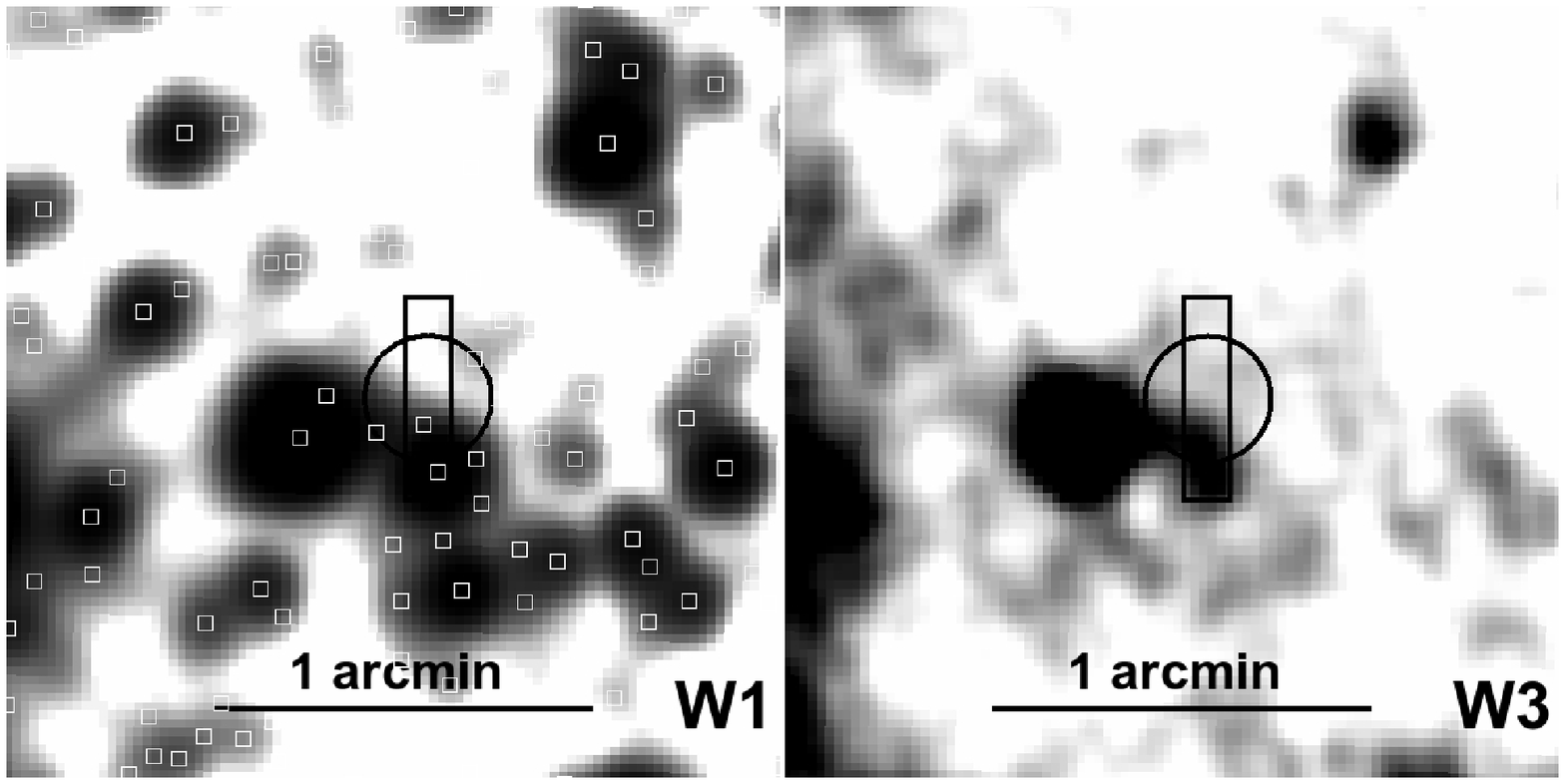}} \\
\caption{{\em WISE} W1 ($3.6$\,$\mu$m) and W3 (12\,$\mu$m) maps around the position of PSR\,J1848--1952. The same field of view is shown as in the {\em Herschel} PACS maps of Figure~\ref{j1848H}. The white boxes in W1 mark position of 2MASS point sources. The circle with a radius of $10\arcsec$ marks the pulsar position, the $7\farcs{4} \times 32\arcsec$ box shows the upper limits of the unknown proper motion together with the radio position uncertainty  (see text).
\label{j1848W}}
\end{center}
\end{figure}
The sky region around PSR\,J1848--1952 is densly populated with infrared sources. Comparing 2MASS, {\em WISE} and our {\em Herschel} observation, it is difficult to pin down common sources. However, we found at least three common sources  between these bands. From the comparison of their positions the expected absolute pointing error of $1\farcs{1}$ for the {\em Herschel} observation seems to be a reasonable estimate. 
There is no source in the {\em Herschel} blue and red band at the position of RX\,J1848--1952 (see Figure~\ref{j1848H}). There is a faint southern source which is inside the box of proper motion upper limits. However this source has a counterpart in other bands, in particular in 2MASS. Therefore, it is not the counterpart of the pulsar.  Using apertures in different source-free regions and at the target position, we determine the $5\sigma$ upper flux limits at the INS position as $F^{\rm blue}_{5\sigma} = 3.6$\,mJy and $F^{\rm red}_{5\sigma} = 5.9$\,mJy.\\ 
The {\em WISE}  W1 and W3 images for the field around PSR\,J1848--1952 are shown in Figure~\ref{j1848W}. 
Except for the southern 2MASS source, there is no {\em WISE} point source deteced in the region of PSR\,J1848--1952.

\section{B. Temperature distribution of dust around a neutron star}
\label{dusttemp}
We know from X-ray and UV observations of XTINSs that their radiation 
can be approximately described as blackbody radiation with temperatures 
$kT_{\rm NS} = 40 - 100$ eV (i.e., $T_{\rm NS} =  0.5 - 1.2 \times 10^6$\,K; see, e.g., \citealt{Haberl2007}).
This radiation heats the dust grains around a XTINS. 
If collisions of a grain with surrounding particles can be neglected,
its temperature $T_\mathrm{g}$ is determined by the balance between 
the radiative heating and cooling
(e.g., \citealt{Backman1993}):
\be
(R_{\rm NS}/r)^2 \int_0^\infty \epsilon^{\rm abs}_\nu (a) \pi B_\nu(T_{\rm NS})\,d\nu = 4 \int_0^\infty \epsilon^{\rm em}_\nu (a) \pi B_\nu(T_\mathrm{g})\,d\nu, 
\label{tempbal}
\ee
where $R_{\rm NS}$ is the stellar radius,
$r$ is the distance of the spherical grain with radius $a$ to the NS, 
$B_\nu(T)$ is the Planck function in dependence on the NS surface temperature, $T_{\rm NS}$, or the grain temperature $T_\mathrm{g}$, $\epsilon^{\rm abs}_\nu (a)$ and $\epsilon^{\rm em}_\nu (a)$ are the grain's absorption and emission efficiencies, which depend on the radiation wavelength (frequency), the grain size,
and the properties of grain material. \\

If the grain radius $a$ is much larger than the wavelength of the heating radiation (e.g., $a \gg 0.03\,\mu$m for $T_{\rm NS}= 0.5\times 10^6$\,K), the absorption efficiency can be
approximated as $\epsilon_\nu^{\rm abs}\approx 1$, i.e., the grains are nearly
perfect absorbers. For the UV and soft X-rays considered here, this approximation is consistent with the detailed calculations by, e.g., \citet{Dwek1996}.
At large distances from the XTINS we expect cold dust grains,
which emit long-wavelength radiation and are inefficient emitters. 
In the Rayleigh limit, 
$\lambda \gg 2\pi a$ (or $\nu\ll c/(2\pi a)$), 
the emission efficiency of a dielectric grain is (e.g., \citealt{Seki1980}): 
\be
\epsilon_\nu^{\rm em} \approx
\frac{8\pi a\nu}{c} {\rm Im}\left[\frac{\varepsilon_\nu -1}{\varepsilon_\nu +2}\right]=
\frac{2\pi a \nu}{c}\frac{24n_\nu k_\nu}{(n_\nu^2 - k_\nu^2+2)^2 + 4 n_\nu^2 k_\nu^2} \ll 1
\label{Rayleigh}
\ee
where $\varepsilon_\nu= (n_\nu + i k_\nu)^2$ is the dielectric constant,
and  $k_\nu$ and $n_\nu$ are the imaginary and real parts of the 
complex refraction coefficient of the grain material.
The quantities $n_\nu$ and $k_\nu$ depend on the grain's chemical composition, 
and their dependence on $\nu$ can be quite complex, including resonances
at some frequencies/wavelengths 
(e.g., at $\lambda \approx 10$ and $\approx 20\,\mu$m for silicate grains; 
see \citealt{Dorschner1995}). 
However, at wavelengths well above the 
resonance wavelength, $k_\nu$ and $n_\nu$ show a universal frequency
dependence:
\be
 n_\nu = n, \qquad\qquad k_\nu = k_0 (\nu/\tilde{\nu}_0) \ll n,\qquad\qquad
{\rm at}\,\,\,\,\, \nu<\tilde{\nu}_0,
\label{longlambda}
\ee
where $\tilde{\nu}_0$ is a frequency below the lowest resonance frequency,
and $k_0$ is a constant that depends on the grain composition.
Using these expressions for $k_\nu$ and $n_\nu$, it is convenient to
parameterize the long-wavelength emission efficiency as follows:
\be
\epsilon_\nu^{\rm em} = \epsilon_0 (\nu/\nu_0)^2
\ee
where
\be
\nu_0 =\left(\frac{\tilde{\nu}_0 c}{2\pi a}\right)^{1/2}, \qquad\qquad
\epsilon_0 = \frac{24 n}{(n^2+2)^2} k_0 ,
\label{nu0-eps0}
\ee
(the last equation takes into account that $k_\nu \ll n$ in this limit).
Substituting the above equations for $\epsilon_\nu^{\rm abs}$ and $\epsilon_\nu^{\rm em}$ into Equation \ref{tempbal}, we obtain
\be
T_{\mathrm g} =\left[\frac{21}{640\pi^3} \frac{L_{\rm NS}}{\sigma_{\rm SB} r^2 \epsilon_0} \left(\frac{h\nu_0}{k}\right)^2\right]^{1/6} =
\left(\frac{21}{640\pi^3} \frac{L_{\rm NS}}{\sigma_{\rm SB}\epsilon_0 r^2}
\frac{h^2c\tilde{\nu}_0}{2\pi k^2 a}\right)^{1/6} ,
\label{Tgrain}
\ee
where $\sigma_{\rm SB}$ is the Stefan Boltzmann constant, $L_{\rm NS}=4\pi R_{\rm NS}^2 \sigma_{\rm SB} T_{\rm NS}^4$ is the star's bolometric luminosity.
These equations are applicable when three conditions are fulfilled:
$\epsilon_\nu <1$, $\nu < \tilde{\nu}_0$, and $\nu < c/2\pi a$, where $\nu\sim 3kT/h$.
The quantities $\epsilon_0$ and ${\nu}_0$ in Equation (\ref{Tgrain}) can
be evaluated from measurements of optical properties of grain material for a chosen $\tilde{\nu}_0$.
Using measurements for amorphous magnesium silicate dust grains as listed in the Jena Database of Optical Constants for Cosmic Dust, we obtain $\epsilon_0 = 0.36$ for $\tilde{\nu}_0 = 6 \times 10^{12}$\,Hz (corresponding to the {\em Herschel} red band wavelength of 160\,$\mu$m). 
This gives the following estimate for the temperature at distance $r$:
\begin{equation}
T_\mathrm{g}= 39 \left(\frac{L_{30}}{r^2_{14} a} \right)^{1/6}\,{\rm K},
\end{equation}
where $L_{30}$ is the bolometric NS luminosity in units of $10^{30}$\,erg\,s$^{-1}$, 
$a$ is the grain radius in $\mu$m, and $r_{14}$ is the distance from the NS in units of $10^{14}$\,cm.
This equation is valid for $T<\min(96$\,K$,2300 a^{-1}_{\mu \mathrm{m}}$\,K$)$.

\end{document}